\documentclass[aps,superscriptaddress,preprintnumbers,showpacs,nofootinbib]{revtex4}
\usepackage{graphicx,wrapfig,amssymb}
\usepackage{color}

\newcommand{\be}{\begin{equation}}
\newcommand{\ee}{\end{equation}}
\newcommand{\ba}{\begin{eqnarray}}
\newcommand{\ea}{\end{eqnarray}}
\newcommand{\la}{\langle}
\newcommand{\ra}{\rangle}
\newcommand{\di}{ {\rm d} }

\newcommand{\asym}[2]
    {\renewcommand{\arraystretch}{0.7}
	\begin{array}{c}{#1}\\ \longleftarrow\!\bullet  \end{array}\!\!\!
	\begin{array}{r}{#2}\\ \longrightarrow \end{array}}

\begin{document}
%===================  TITLE, AUTHORS, AFFILIATIONS ===================
\newcommand*{\Dubna}{Joint Institute for Nuclear Research, Dubna, 
141980 Russia}\affiliation{\Dubna}
\newcommand*{\Bochum}{Institut f{\"u}r Theoretische Physik II, 
Ruhr-Universit{\"a}t Bochum, D-44780 Bochum, Germany}\affiliation{\Bochum}

\title{Collins effect in semi-inclusive deeply inelastic scattering 
	and in \boldmath $e^+e^-$-annihilation}
\author{A.~V.~Efremov}\affiliation{\Dubna}
\author{K.~Goeke}\affiliation{\Bochum}
\author{P.~Schweitzer}\affiliation{\Bochum}
\date{May 2006}
%===================  PREPRINT NUMBER, JOURNAL =======================
\preprint{RUB-TPII-02/06} % hep-ph/0603054 (v2)
%===================  ABSTRACT =======================================
\begin{abstract}
  The Collins fragmentation function is extracted from HERMES data on
  azimuthal single spin asymmetries in semi-inclusive deeply inelastic 
  scattering, and BELLE data on azimuthal asymmetries in $e^+e^-$-annihilations.
  A Gaussian model is assumed for the distribution of transverse parton momenta 
  and predictions are used from the chiral quark-soliton model for the transversity 
  distribution function.
  We find that the HERMES and BELLE data yield a consistent picture of the
  Collins fragmentation function which is compatible with COMPASS data and 
  the information previously obtained from an analysis of DELPHI data. 
  Estimates for future experiments are made.
\end{abstract}
 % Hadron-induced inclusive production with identified leptons, 
                % photons, or other nonhadronic particles (energy > 10 GeV)
\maketitle
%===================  SECTION 1: INTRODUCTION ========================
\section{Introduction}
\label{Sec-1:introduction}

The chirally odd Collins fragmentation function $H_1^\perp$ describes a possible 
left-right-asymmetry in the fragmentation of transversely polarized quarks
\cite{Efremov:1992pe,Collins:1992kk,Collins:1993kq}, the effects of which can be 
observed in two ways. 
In $e^+e^-$ annihilations it may give rise to specific azimuthal asymmetries 
\cite{Boer:1997mf,Boer:1997qn} where one can access $H_1^{\perp q}H_1^{\perp\bar q}$
adequately weighted with electric or electro-weak charges. 
In semi-inclusive deeply inelastic scattering (SIDIS) with a transversely (or 
longitudinally) polarized target it may give rise to azimuthal single spin asymmetries
(SSA) \cite{Mulders:1995dh,Boer:1997nt} where it enters in combination with the 
transversity distribution function $h_1^q(x)$ \cite{Ralston:1979ys,Barone:2001sp}
(or other chirally-odd distribution functions).
For these observables transverse parton momenta play a crucial role requiring 
a careful generalization \cite{Collins:1981uk,Ji:2004wu,Collins:2004nx,Metz:2002iz} 
of standard (``collinear'') factorization theorems \cite{Mueller:1989hs}.

% NEW:
% The first in principle unambiguously interpretable experimental indications for the
% Collins effect were obtained from a study of SMC data from SIDIS with transversely
% polarized targets \cite{Bravar:1999rq} and from a study of DELPHI data on charged
% hadron production in $e^+e^-$ annihilations at the $Z^0$-pole \cite{Efremov:1998vd}.
The first evidence for the Collins fragmentation function was looked for in a study 
of SMC data from SIDIS with transversely polarized targets \cite{Bravar:1999rq}, 
and in a study of DELPHI data on charged hadron production in $e^+e^-$ 
annihilations at the $Z^0$-pole \cite{Efremov:1998vd}.
% END NEW.
More recently the HERMES Collaboration has published (and shown further preliminary)
data on the Collins SSA in SIDIS from a proton target giving the first unambiguous 
evidence that $H_1^\perp$ and $h_1^a(x)$ are non-zero 
\cite{HERMES-new,Airapetian:2004tw,Diefenthaler:2005gx},
while in the COMPASS experiment the Collins effect from a deuteron target
was found compatible with zero within error bars \cite{Alexakhin:2005iw}.
Finally, the BELLE collaboration presented data on sizeable azimuthal asymmetries 
in $e^+e^-$ annihilations at a center of mass energy of $60\,{\rm MeV}$ below the 
$\Upsilon$-resonance \cite{Abe:2005zx}. 

One question which immediately arises is: 
{\sl Are all these data from different SIDIS and $e^+e^-$ experiments compatible,
i.e.\  are they all indeed due to the same effect, namely the Collins effect?}

In order to answer this question one may try to fit the HERMES 
\cite{Airapetian:2004tw,Diefenthaler:2005gx}, COMPASS \cite{Alexakhin:2005iw}
and BELLE \cite{Abe:2005zx} data simultaneously. 
While worthwhile trying \cite{Anselmino:2005kn} such a procedure has also 
disadvantages. When using SIDIS data the obtained fit for the Collins fragmentation
function is inevitably biased by some model assumption on the transversity 
distribution, and the --- presently not fully understood --- scale dependence 
of $H_1^\perp$ is neglected.

For these reasons we shall follow a different strategy and extract $H_1^\perp$
separately from HERMES \cite{Diefenthaler:2005gx} and BELLE \cite{Abe:2005zx} data, 
and compare then the results to each other and to other experiments. 

% NEW:
% By focussing on adequately defined ratios of $H_1^\perp$ to $D_1$
% (``analyzing powers'') we observe that the DELPHI \cite{Efremov:1998vd},
% HERMES \cite{HERMES-new,Airapetian:2004tw,Diefenthaler:2005gx}, COMPASS
% \cite{Alexakhin:2005iw} and BELLE \cite{Abe:2005zx} data are compatible with
% each other.
By focussing on adequately defined ratios of $H_1^\perp$ to $D_1$ 
(``analyzing powers'') we observe that 
the indications from DELPHI \cite{Efremov:1998vd} and the data from
HERMES \cite{HERMES-new,Airapetian:2004tw,Diefenthaler:2005gx}, COMPASS 
\cite{Alexakhin:2005iw} and BELLE \cite{Abe:2005zx} are compatible with each other. 
% END NEW.
Experience with spin observables where radiative and other corrections are known, 
see e.g.\  \cite{Ratcliffe:1982yj,Kotikov:1997df}, suggests that such analyzing 
powers could be less scale-dependent than the respective absolute cross sections.

For the analysis of the HERMES data we use \cite{talk-at-SIR05} predictions 
for $h_1^a(x)$ from the chiral quark-soliton model \cite{Schweitzer:2001sr}
and assume a Gaussian distribution of transverse parton momenta.
Our results are in agreement with the first extraction of $H_1^\perp$ from the 
HERMES data \cite{Vogelsang:2005cs}, where different models were assumed.\footnote{
	\label{footnote-1}
	Previously, the 
	preliminary result \cite{Bravar:1999rq} and longitudinal target
	SSA \cite{Avakian:1999rr,Airapetian:2002mf,Avakian:2003pk,Airapetian:2005jc}
	data were used to extract $H_1^\perp$ from SIDIS 
	\cite{Anselmino:2000mb,Efremov:2001cz}. However, in these studies twist-3 
	effects \cite{Afanasev:2003ze} were not or only partially considered.
	Still earlier attempts to learn about the Collins effect from SSA 
	in the hadronic processes $\bar pp^\uparrow$ or $pp^\uparrow\to\pi X$ 
	\cite{Adams:1991rw} were made in \cite{Anselmino:1999pw}, but recent studies 
	indicate that the Collins effect is not likely to be the dominant 
	source of SSA in these processes \cite{Anselmino:2004ky} as it
	was assumed in \cite{Anselmino:1999pw}.}
On the basis of a comparison to Ref.~\cite{Vogelsang:2005cs} we discuss which 
of the observations made here and in \cite{Vogelsang:2005cs} are robust, and 
which are likely to be model-dependent.

We shall furthermore present some estimates for future experiments in SIDIS 
and at $e^+e^-$ colliders. The comparison of these estimates to the outcome 
of the experiments will help to solidify our understanding of the Collins 
fragmentation function and the transversity distribution.

The SIDIS data from HERMES and COMPASS 
\cite{HERMES-new,Airapetian:2004tw,Diefenthaler:2005gx,Alexakhin:2005iw} also 
provide the first unambiguous evidence for the equally interesting  Sivers 
effect \cite{Sivers:1989cc,Brodsky:2002cx,Collins:2002kn,Belitsky:2002sm}. 
We refer the reader to Refs.~\cite{Vogelsang:2005cs,Efremov:2004tp,Anselmino:2005nn,Anselmino:2005ea,Collins:2005ie,Collins:2005wb,Anselmino:2005an}
for studies of this effect.

%===================  SECTION 2: COLLINS IN SIDIS ====================
\section{Collins effect in SIDIS}
\label{Sec-2:Collins-effect-in-SIDIS}

For reasons which will become clear in the sequel we start our study with the 
Collins effect in SIDIS --- although this case is more involved because here in 
addition to $H_1^\perp$ also the unknown transversity distribution $h_1^a(x)$ enters.
For the latter we use here predictions from the chiral quark-soliton model 
\cite{Schweitzer:2001sr} which provides a consistent and successful 
\cite{Diakonov:1987ty,Christov:1995vm} field theoretic description of the nucleon. 
The distribution functions $f_1^a(x)$ and $g_1^a(x)$ computed in this model at 
a low normalization point \cite{Diakonov:1996sr} agree to within (10-30)$\%$ 
with parameterizations \cite{Gluck:1998xa}. One may assume the same accuracy 
for the model predictions for transversity, which are inbetween the popular 
working assumptions, namely that at a low scale $h_1^q(x)$ saturates the Soffer 
bound $|h_1^a(x)|\le\frac12(f_1^a+g_1^a)(x)$ \cite{Soffer:1994ww} or is equal 
to $g_1^q(x)$.

The way the HERMES \cite{Airapetian:2004tw,Diefenthaler:2005gx} (and BELLE 
\cite{Abe:2005zx}) data were analyzed requires to assume also a model for 
transverse parton momenta. We assume the distributions of transverse parton momenta 
in the transversity distribution and transverse hadron momenta in the Collins 
fragmentation function to be Gaussian, and take the respective widths 
$\la{\bf p}_{h_1}^2\ra$ and $\la{\bf K}_{H_1}^2\ra$ to be flavour and 
$x$- or $z$-independent
\ba\label{Eq:Gauss-ansatz}
   	h_1^a(x,{\bf p}_T^2) & \equiv & h_1^a(x)\; 
	\frac{\exp(-{\bf p}_T^2/\la{\bf p}^2_{h_1}\ra)}{\pi\;\la{\bf p}^2_{h_1}\ra}
	\;,\nonumber\\
	H_1^{\perp a}(z,{\bf K}_T^2) & \equiv & H_1^{\perp a}(z)\;
	\frac{\exp(-{\bf K}_T^2/\la{\bf K}^2_{H_1}\ra)}{\pi\;\la{\bf K}^2_{H_1}\ra}
	\;.\ea
The Gauss Ansatz can be considered at best as a crude approximation to 
the true distribution of transverse momenta \cite{Collins:2003fm}.
For example, it does not yield a $1/P_{h\perp}$ power-like suppression of the 
Collins SSA expected at large transverse hadron momenta \cite{Collins:1992kk}. 
However, this Ansatz gives a satisfactory effective description in several hard 
reactions, provided the transverse momenta are much smaller than the 
hard scale of the process \cite{D'Alesio:2004up}. This is the situation in the 
HERMES experiment where the mean transverse hadron momenta
$\la P_{h\perp}\ra\sim0.4\,{\rm GeV}\ll \sqrt{\la Q^2\ra}\sim 1.5\,{\rm GeV}$.

%------ BEGIN FIGURE 1: Kinematics of SIDIS ---------------------------
\begin{wrapfigure}{RD}{6cm}
	\centering
	\vspace{0.5cm}
        \includegraphics[width=6cm]{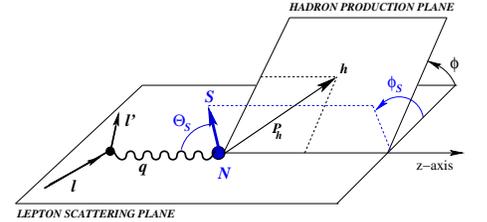}
        \caption{\label{fig1-processes-kinematics}
	Kinematics of the SIDIS process $lp\to l^\prime h X$
	and the definitions of azimuthal angles in the lab frame.	
	The target polarization vector is transverse with respect to the beam.}
\end{wrapfigure}
%------ END FIGURE 1 -------------------------------------------------
%
In fact, the HERMES data on $\la P_{h\perp}\ra$ can be described 
satisfactorily assuming for $f_1^a(x,{\bf p}_T^2)$ and $D_1^a(z,{\bf K}_T^2)$
Ans\"atze analog to (\ref{Eq:Gauss-ansatz}) with \cite{Collins:2005ie}
\be\label{Eq:Gauss-width-unp}
	\la{\bf p}^2_{f_1}\ra=0.33\,{\rm GeV}^2\;\;\;\mbox{and}\;\;\;
	\la{\bf K}^2_{D_1}\ra=0.16\,{\rm GeV}^2\,.\ee
Noteworthy these numbers are in good agreement with results 
\cite{Anselmino:2005nn} inferred from EMC data \cite{Arneodo:1986cf}
on the so-called Cahn-effect \cite{Cahn:1978se}. All that can be said about the
Gaussian widths of the transversity distribution and Collins function is that
positivity conditions require \cite{Bacchetta:1999kz}
\ba\label{Eq:positivity}
	|h_1^a(x,{\bf p}^2_T)| \; \le \; f_1^a(x,{\bf p}^2_T) 
	& \Rightarrow &
	\la{\bf p}^2_{f_1}\ra \le \la{\bf p}^2_{h_1}\ra \nonumber \\
	\frac{|{\bf K}_T|}{zm_\pi}H_1^{\perp a}(z,{\bf K}_T^2)\le D_1^a(z,{\bf K}_T^2)
	& \Rightarrow &
	\la{\bf K}^2_{H_1}\ra < \la{\bf K}^2_{D_1}\ra \;
\ea
which are necessary conditions (but not sufficient for $H_1^\perp$, see 
\cite{Collins:2005ie} for the discussion of the analog case of the Sivers function).

In the Gaussian model (\ref{Eq:Gauss-ansatz}) the Collins SSA as measured by HERMES 
\cite{Airapetian:2004tw,Diefenthaler:2005gx} is given by (we neglect throughout the
soft factors \cite{Ji:2004wu,Collins:2004nx}) 
\ba\label{Eq:AUT-Collins-1}
&&	A_{UT}^{\sin(\phi+\phi_S)} \equiv 2\la \sin(\phi+\phi_S)\ra
	\nonumber\\
&&	= 2\;\frac{
	\sum_a e_a^2\, x h_1^a(x)\,B_{\rm Gauss} H_1^{\perp(1/2)a}(z)}{
	\sum_a e_a^2\,x f_1^a(x)\,D_1^{a}(z)} \;,
\ea
where $x=Q^2/(2P\cdot q)$, $z=P\cdot P_h / P\cdot q$, $Q^2=- q^2$ and
$P$ is the target momentum. Other momenta and azimuthal angles are defined 
in Fig.~\ref{fig1-processes-kinematics}.
The $(1/2)$-moment of the Collins function is defined as
and satisfies the inequality 
\be\label{Eq:H1perp-1/2mom-and-bound}
	H_1^{\perp (1/2)a}(z)= \int\!\di^2{\bf K}_T\;\frac{|{\bf K}_T|}{2zm_\pi}
	H_1^{\perp      a}(z, {\bf K}_T) \le \frac12\;D_1^a(z)\;,
\ee
which follows from (\ref{Eq:positivity}), see \cite{Bacchetta:1999kz}.
In Eq.~(\ref{Eq:AUT-Collins-1}) the dependence on the Gaussian model is 
contained in the factor
\be\label{Eq:B_Gauss}
	B_{\rm Gauss}(z) = 
	\frac{1}{\sqrt{1+z^2\;\la{\bf p}_{h_1}^2\ra/\la{\bf K}^2_{H_1}\ra}\;}\;.
\ee
Within the Gaussian model Eq.~(\ref{Eq:AUT-Collins-1}) is just one convenient way 
of writing the result. Equivalently one could write 
\be\label{Eq:a_Gauss}
	B_{\rm Gauss} H_1^{\perp (1/2)a}(z) = 
	a_{\rm Gauss} H_1^{\perp (1)a}(z) \;,\;\;\;
	a_{\rm Gauss} = \frac{\sqrt{\pi}\,m_\pi}
	{2\sqrt{\la{\bf p}_{h_1}^2\ra+\la{\bf K}_{H_1}^2\ra/z^2\,}}\;.
\ee
Here the $(1)$-transverse moment of the Collins function is defined as in 
Eq.~(\ref{Eq:H1perp-1/2mom-and-bound}) but with an additional power of 
$|{\bf K}_T|/(zm_\pi)$ in the weight, and $a_{\rm Gauss}$ is defined in 
complete analogy to the notation of Ref.~\cite{Collins:2005ie}.

Had the Collins SSA been analyzed as 
$\la\sin(\phi+\phi_S)P_{h\perp}/(z m_\pi)\ra$, i.e. with an additional power 
of transverse hadron momentum $P_{h\perp}$ in the weight, the result would 
be given by an expression analog to Eq.~(\ref{Eq:AUT-Collins-1}) but with 
$B_{\rm Gauss}\to 1$ and $H_1^{\perp(1/2)}$ replaced by $H_1^{\perp(1)}$ 
{\sl independently} of any model of transverse parton momenta \cite{Boer:1997nt}. 
It was argued that adequately weighted SSA might be less sensitive to Sudakov 
suppression effects \cite{Boer:2001he}. Preliminary HERMES data analyzed in 
this way were presented in \cite{HERMES-new}. 

By defining  
$A_\pi = A_{UT,\pi}^{\sin(\phi+\phi_S)}\sum_a e_a^2\,x f_1^a\,D_1^{a/\pi}$
and using HERMES data \cite{Airapetian:2004tw,Diefenthaler:2005gx} on $\pi^\pm$ 
we can rewrite (\ref{Eq:AUT-Collins-1}) as
\be\label{Eq:AUT-Collins-2}
	\left(\matrix{A_{\pi^+} \cr A_{\pi^-}}\right) = 
	\left(\matrix{
	\frac49xh_1^u+\frac19xh_1^{\bar d} & \frac19xh_1^d+\frac49xh_1^{\bar u} \cr
	\frac19xh_1^d+\frac49xh_1^{\bar u} & \frac49xh_1^u+\frac19xh_1^{\bar d}}
	\right) \cdot \left(\matrix{
	2B_{\rm Gauss}H_1^{\perp(1/2)\rm fav} \cr
	2B_{\rm Gauss}H_1^{\perp(1/2)\rm unf} }\right)\;,
\ee
which --- with our chosen model for $h_1^a(x)$ --- can be inverted to give 
unambiguous results for $B_{\rm Gauss}H_1^{\perp(1/2)a}$. The favoured 
and unfavoured Collins fragmentation functions are defined as, schematically:
\ba\label{Eq:H-fav-unf}
    &&	H_1^{\rm fav}=
	H_1^{u/\pi^+}=H_1^{d/\pi^-}=H_1^{\bar u/\pi^-}=H_1^{\bar d/\pi^+}\;,\nonumber\\
    &&	H_1^{\rm unf}=
	H_1^{u/\pi^-}=H_1^{d/\pi^+}=H_1^{\bar u/\pi^+}=H_1^{\bar d/\pi^-}\;.
\ea

%------ BEGIN FIGURE 2: B_Gauss H1perp(1/2) ---------------------------
        \begin{wrapfigure}{RD}{6.5cm}
	\vspace{-0.5cm}
        %\centering
        \includegraphics[width=6.5cm]{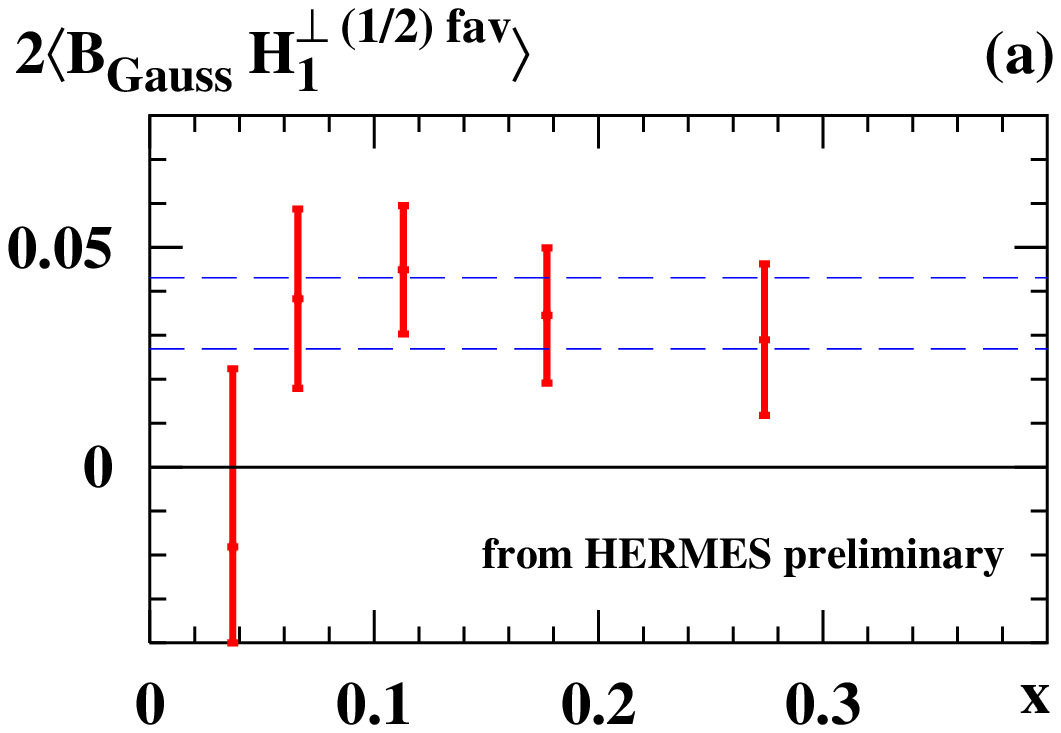}
	\vspace{0.1cm}

        \includegraphics[width=6.5cm]{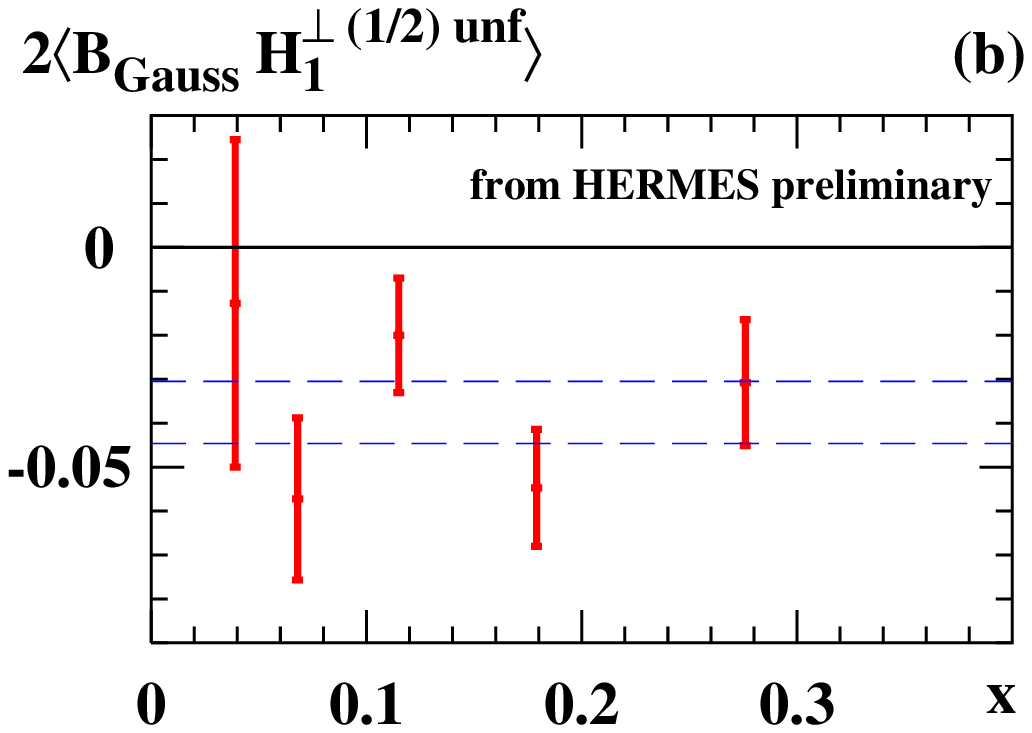}
        \caption{\label{fig2:B-Gauss-H1perp12} 
	The quantity $2B_{\rm Gauss}H_1^{\perp(1/2)a}$ averaged over $z$,
	i.e.\  practically the weight of $h_1^a(x)$ in the Collins SSA 
	$A_{UT}^{\sin(\phi+\phi_S)}(x)$ in Eq.~(\ref{Eq:AUT-Collins-1}), 
	vs.\ $x$ as extracted from the preliminary HERMES data 
	\cite{Diefenthaler:2005gx}. This quantity does not show any
	significant $x$-dependence --- as expected, see text.}
	%\vspace{-0.8cm}	
        \end{wrapfigure}
%------ END FIGURE 2 --------------------------------------------------
\noindent
Here --- in the study of SIDIS data --- we neglect the effects of strange and 
heavier flavours, which is justified at the present stage because the corresponding 
distribution functions are rather small. 

As a first step we extract in this way from the HERMES preliminary data 
\cite{Diefenthaler:2005gx} on the $x$-dependence of the Collins SSA the quantity 
$\la 2B_{\rm Gauss}H_1^{\perp(1/2)a}\ra$ which is averaged over $z$ within the
HERMES cuts $0.2\le z \le 0.7$. Using for $h_1^a(x)$ the chiral quark soliton model 
\cite{Schweitzer:2001sr} and for $f_1^a(x)$ and $D_1^a(z)$ the LO-parameterizations 
\cite{Gluck:1998xa,Kretzer:2001pz} at $\la Q^2\ra=2.5\,{\rm GeV}^2$ which is the 
average scale in the HERMES experiment, we obtain the results shown in 
Fig.~\ref{fig2:B-Gauss-H1perp12}. 
As the SSA for $\pi^+$ and $\pi^-$ are independent observables the statistical 
errors shown in Fig.~\ref{fig2:B-Gauss-H1perp12} are not correlated.

The $\la 2B_{\rm Gauss}H_1^{\perp(1/2)a}\ra$ are expected to be $x$-independent
--- provided the used models, the Gaussian Ansatz (\ref{Eq:Gauss-ansatz}) 
and the model \cite{Schweitzer:2001sr} for $h_1^a(x)$, are appropriate. 
In fact, the extracted $\la 2B_{\rm Gauss}H_1^{\perp(1/2)a}\ra$ are compatible 
with this expectation and can be fitted respectively to constants with 
reasonable $\chi^2$ per degree of freedom ($\chi^2_{\rm dof}$)
\ba\label{Eq:B-Gauss-H1perp12-fav}
   &&	\la 2B_{\rm Gauss}H_1^{\perp(1/2)\rm fav}\ra =\;\;\;(3.5\pm 0.8)\% \;, \;\;\; 
	\chi^2_{\rm dof}=0.6 \;,\\
   \label{Eq:B-Gauss-H1perp12-unf}
   &&	\la 2B_{\rm Gauss}H_1^{\perp(1/2)\rm unf}\ra = -(3.8\pm 0.7)\% \;, \;\;\; 
	\chi^2_{\rm dof}=1.3 \;.
\ea
These constants are indicated in Fig.~\ref{fig2:B-Gauss-H1perp12}. The 
published HERMES data which have a lower statistics \cite{Airapetian:2004tw} 
yield similar values but with larger uncertainties and worse $\chi^2_{\rm dof}$.

Let us remark that at small $x$ a deviation from the 
expected behaviour $\la 2B_{\rm Gauss}H_1^{\perp(1/2)a}(x)\ra=$ constant would not 
be surprizing for two reasons. 
First, the description of distribution functions in the small-$x$ region 
$x\lesssim 0.05$ is beyond the range of applicability of the chiral quark-soliton 
model \cite{Diakonov:1996sr}. Second, in the lowest $x$-bin of the HERMES experiment 
$\la Q^2\ra = 1.3\,{\rm GeV}$ \cite{Airapetian:2005jc}  could be at the 
edge of the applicability of the factorization approach.

However, the lack of a noticeable $x$-dependence of 
$\la 2B_{\rm Gauss}H_1^{\perp(1/2)a}\ra$ indicated in Fig.~\ref{fig2:B-Gauss-H1perp12} 
gives certain confidence that the model \cite{Schweitzer:2001sr} for $h_1^a(x)$ used 
here is --- considering the accuracy of the data 
\cite{Airapetian:2004tw,Diefenthaler:2005gx} --- reasonable. The results in 
Eqs.~(\ref{Eq:B-Gauss-H1perp12-fav},~\ref{Eq:B-Gauss-H1perp12-unf}) are, of course,
model-dependent but Fig.~\ref{fig2:B-Gauss-H1perp12} indicates that the 
systematic uncertainty due to model-dependence is less dominant than the statistical 
error in (\ref{Eq:B-Gauss-H1perp12-fav},~\ref{Eq:B-Gauss-H1perp12-unf}). 

We also remark that the results in Fig.~\ref{fig2:B-Gauss-H1perp12} and 
Eqs.~(\ref{Eq:B-Gauss-H1perp12-fav},~\ref{Eq:B-Gauss-H1perp12-unf}) are not 
specific to the Gaussian Ansatz: Any analysis of the data \cite{Diefenthaler:2005gx}
in which the factorized Ansatz $h_1^a(x,{\bf p}_T^2)=h_1^a(x)\,G({\bf p}_T^2)$ 
is assumed with $G({\bf p}_T^2)$ some function of ${\bf p}_T^2$ (and $h_1^a(x)$ 
is taken from \cite{Schweitzer:2001sr}) would yield the same result 
(\ref{Eq:B-Gauss-H1perp12-fav},~\ref{Eq:B-Gauss-H1perp12-unf}).
(Then, however, this $z$-averaged quantity would be 
given by some different theoretical expressions.)

The results (\ref{Eq:B-Gauss-H1perp12-fav},~\ref{Eq:B-Gauss-H1perp12-unf}) 
are unexpected from the point of view of studies 
\cite{Efremov:2001cz,DeSanctis:2000fh,Efremov:2001ia,Ma:2002ns,Efremov:2004hz} 
of longitudinal SSA \cite{Avakian:1999rr,Airapetian:2002mf}. In order to understand 
these data it was sufficient to {\sl assume} favoured flavour fragmentation only and to 
neglect $H_1^{\perp\rm unf}$ completely, see \cite{Efremov:2004hz} for a review. 
Later it became clear that these SSA are dominated by subleading twist effects 
\cite{Airapetian:2005jc}, and are theoretically more difficult to describe
\cite{Afanasev:2003ze}. Moreover, we now learn that the unfavoured fragmentation 
in the Collins function cannot be neglected. Instead, in order to explain the HERMES 
data \cite{Airapetian:2004tw,Diefenthaler:2005gx} the favoured and unfavoured Collins 
fragmentation functions must be of similar magnitude and opposite sign.
The string fragmentation picture provides a 
qualitative understanding of this behaviour \cite{Artru:1995bh} as do 
quark-hadron-duality motivated phenomenological considerations \cite{Vogelsang:2005cs}.

The different behaviour of the Collins function compared to the unfavoured
fragmentation function becomes more evident by considering the analyzing powers
defined as
\be\label{Eq:Apower-HERMES}
	\frac{\la 2 B_{\rm Gauss}H_1^{\perp(1/2)\rm fav}\ra}{\la D_1^{\rm fav}\ra}
	\biggr|_{\rm HERMES}\; = (7.2\pm 1.7)\% \;, \;\;\;\; 
  	\frac{\la 2 B_{\rm Gauss}H_1^{\perp(1/2)\rm unf}\ra}{\la D_1^{\rm unf}\ra} 
	\biggr|_{\rm HERMES}\; = -(14.2\pm 2.7)\% \;.
\ee
The positivity bound (\ref{Eq:H1perp-1/2mom-and-bound}) and the fact that 
$B_{\rm Gauss}<1$, see Eq.~(\ref{Eq:B_Gauss}), require the absolute values of the 
above numbers to be smaller than unity which is the case. Thus, the extracted results 
satisfy positivity.

%------ BEGIN FIGURE 3: AUT(x) AT HERMES AND COMPASS ------------------
%
\begin{figure}
\begin{tabular}{cccc}
\hspace{-0.5cm}
\includegraphics[width=1.75in]{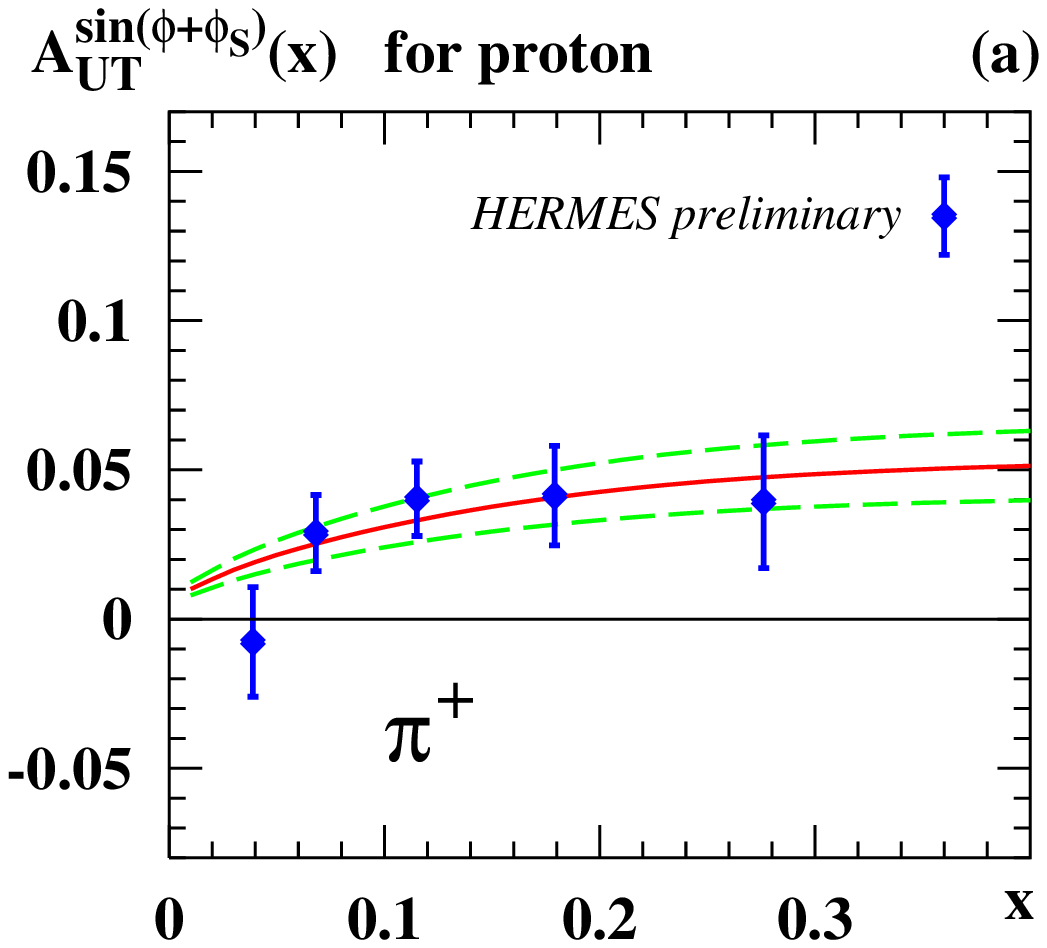}&
\includegraphics[width=1.75in]{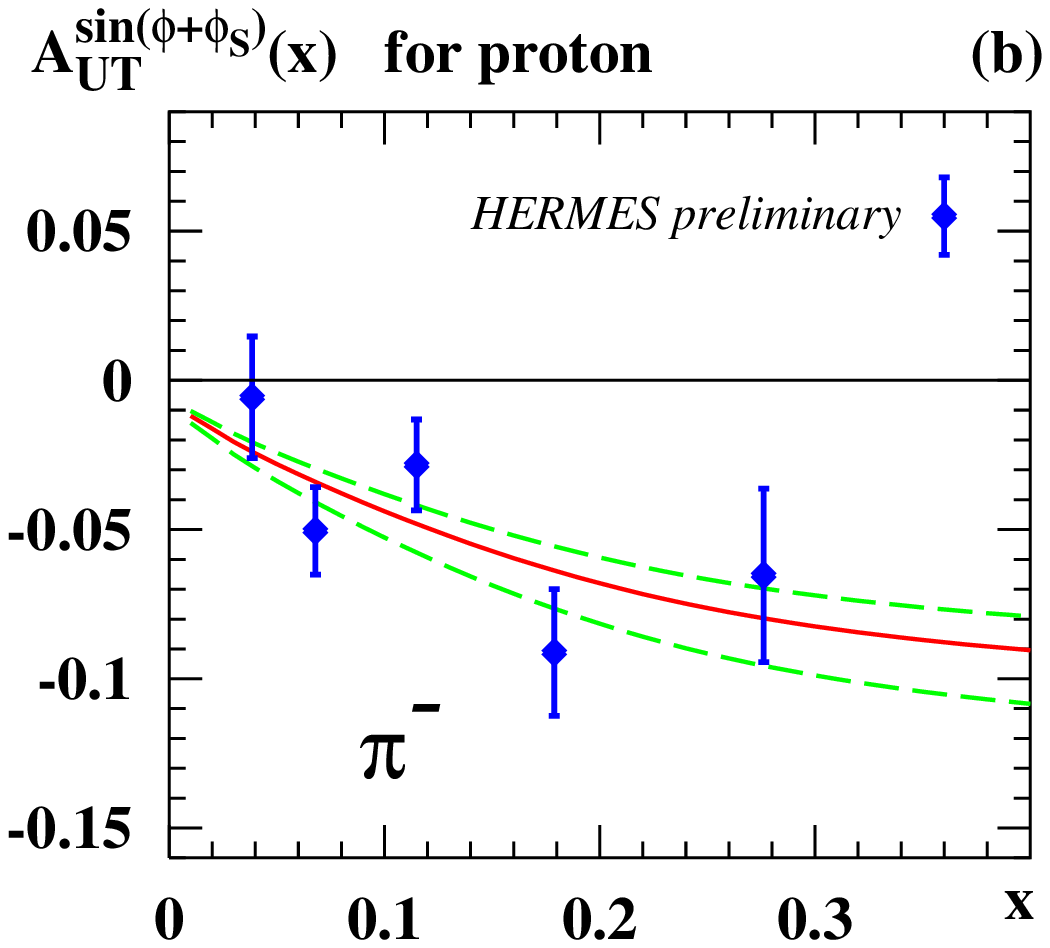}&
\includegraphics[width=1.75in]{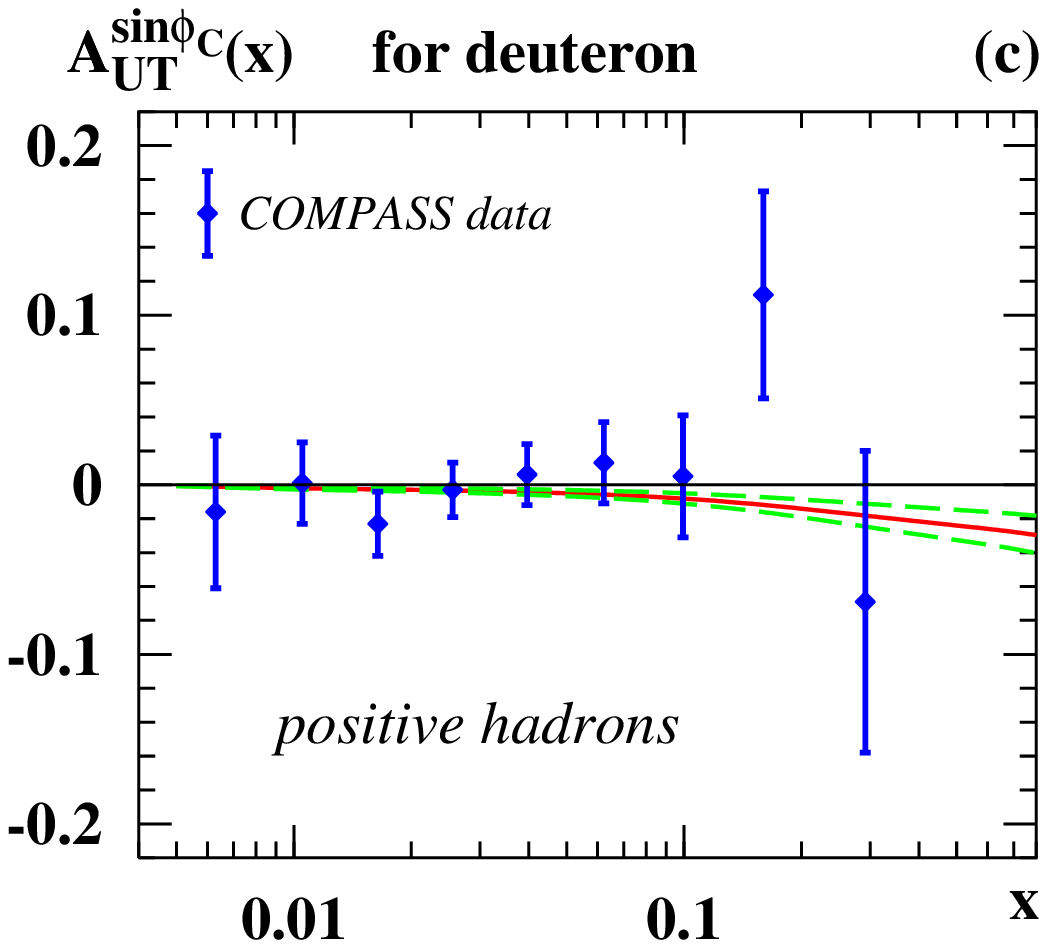}&
\includegraphics[width=1.75in]{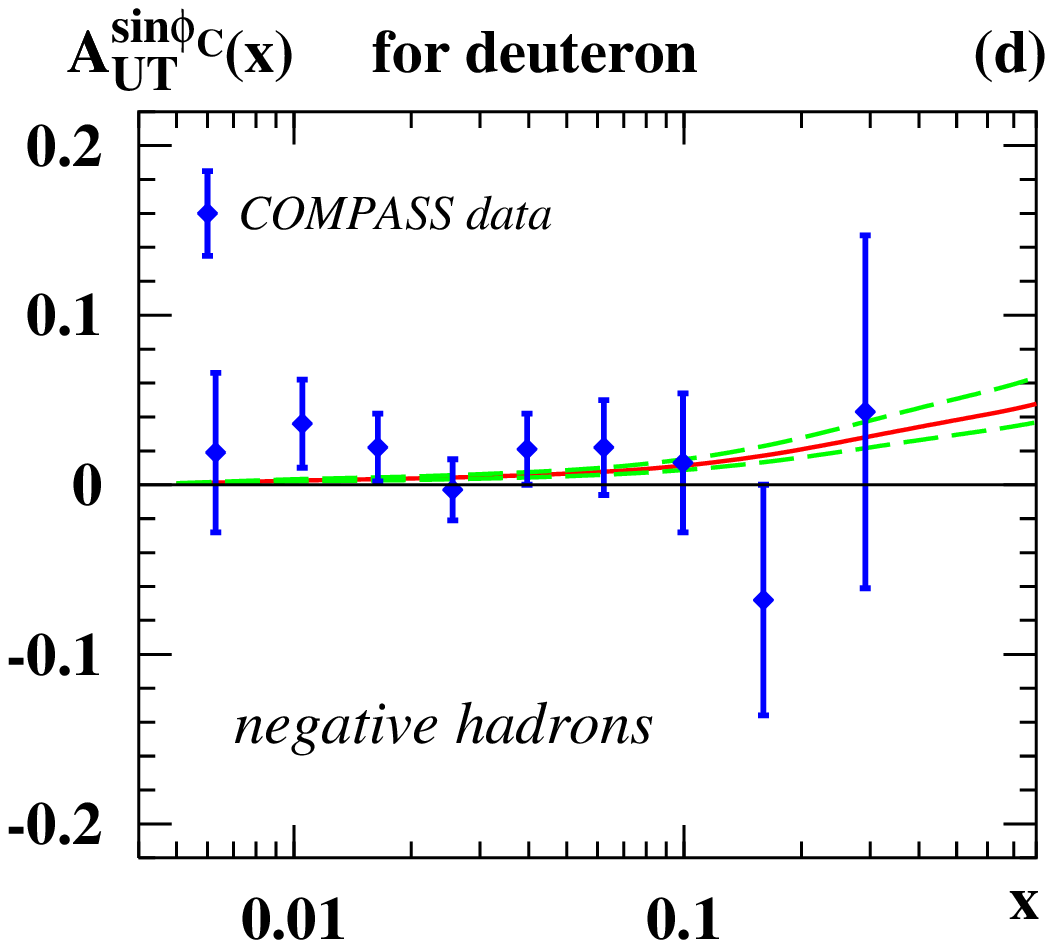}
\end{tabular}
\caption{\label{Fig3:AUT-x}
	The Collins SSA $A_{UT}^{\sin(\phi+\phi_S)}(x)$ as function of $x$.
	The preliminary HERMES data are from \cite{Diefenthaler:2005gx},
	the COMPASS data are from \cite{Alexakhin:2005iw}. 
	The theoretical curves are based on the fit in 
	Eqs.~(\ref{Eq:B-Gauss-H1perp12-fav},~\ref{Eq:B-Gauss-H1perp12-unf})
	and predictions for the transversity distribution from the chiral
	quark-soliton model \cite{Schweitzer:2001sr}. Notice that different 
	sign conventions are used in \cite{Diefenthaler:2005gx,Alexakhin:2005iw}: 
	$A_{UT}^{\sin\phi_C}(x)=-A_{UT}^{\sin(\phi+\phi_S)}(x)$.}
\end{figure}
%
%------ END FIGURE 3 --------------------------------------------------

Figs.~\ref{Fig3:AUT-x}a and b shows how the fit 
(\ref{Eq:B-Gauss-H1perp12-fav},~\ref{Eq:B-Gauss-H1perp12-unf}) describes the 
preliminary HERMES data \cite{Diefenthaler:2005gx} on the Collins SSA from 
the proton target. As can be seen from Figs.~\ref{Fig3:AUT-x}c and d this 
fit is also in agreement with the COMPASS data which show a compatible with
zero Collins effect from a deuteron target \cite{Alexakhin:2005iw}.

In principle, one could extract in an analog way also the $z$-dependence of 
the product $B_{\rm Gauss}H_1^{\perp(1/2)a}$. However, we refrain from doing 
this here for the following reasons. First, due to the appearance of the unknown 
Gaussian widths in $B_{\rm Gauss}$, see Eq.~(\ref{Eq:B_Gauss}), we could not 
conclude anything on the $z$-dependence of $H_1^{\perp(1/2)a}$. Second, the
study of the $z$-dependence requires to average over $x$ within 
the HERMES cuts $0.023\le x\le 0.4$ which includes small $x$ where the model 
\cite{Schweitzer:2001sr} is not applicable. We postpone the study of the 
$z$-dependence of the Collins SSA for Sec.~\ref{SecX:z-dependence-of-SIDIS-data}.

%\newpage
%===================  SECTION 3: COLLINS AT BELLE ====================
\section{Collins effect in $e^+e^-$ annihilation at BELLE}
\label{Sec-3:Collins-at-BELLE}

%------ BEGIN FIGURE 4: Kinematics of e+e- ----------------------------
\begin{wrapfigure}{RD}{6.2cm}
	\centering
        \includegraphics[width=6.2cm]{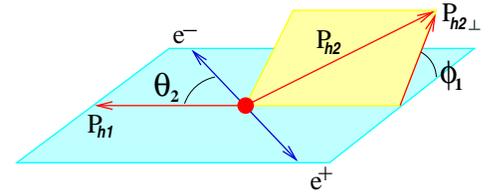}
        \caption{\label{fig4-kinematics-e+e-}
	Kinematics of the process $e^+e^-\to h_1 h_2 X$
	and the definitions of azimuthal angles in the $e^+e^-$ rest frame.}
\end{wrapfigure}
%------ END FIGURE 4 -------------------------------------------------
%

In the process $e^+e^-\to h_1h_2 X$ where the two observed hadrons belong
to opposite jets the Collins effect gives rise to a specific azimuthal
distribution of hadron $h_1$ along the axis defined by hadron $h_2$ 
in the $e^+e^-$ center of mass frame \cite{Boer:1997mf,Boer:1997qn}, 
see Fig.~\ref{fig4-kinematics-e+e-}.
Assuming the Gaussian model this azimuthal distribution is given by
\ba\label{Eq:azim-distr-e+e-}
	\frac{\di^4\sigma(e^+e^-\!\!\to h_1h_2X)}
             {\di\cos\theta_2\di z_1\di z_2\di\phi_1}
	= \frac{3\alpha^2}{Q^2}\biggl[
	(1+\cos^2\theta_2)\sum_ae_a^2D_1^a(z_1)D_1^{\bar a}(z_2) \nonumber\\
	+ \cos(2\phi_1)\,\sin^2\theta_2\,C_{\rm Gauss}\!\!
	\sum_ae_a^2H_1^{\perp(1/2)a}(z_1)H_1^{\perp(1/2)\bar a}(z_2)
	\biggr]
\ea
where $Q^2$ is the center of mass energy square of the lepton pair
and $z_i=2E_{h_i}/Q$ while $C_{\rm Gauss}$ is defined as
\be
	C_{\rm Gauss}(z_1,z_2)=\frac{16}{\pi}\;\frac{z_1z_2}{z_1^2+z_2^2}\;.
\ee
In principle, one can rewrite the result in (\ref{Eq:azim-distr-e+e-}) within the 
Gauss model also in terms of $H_1^\perp$ or $H_1^{\perp(1)}$ with some different 
functions instead of $C_{\rm Gauss}$. 

Experimentally it is convenient to normalize the expression (\ref{Eq:azim-distr-e+e-})
with respect to its $\phi_1$-average and to define the following observable  
\cite{Abe:2005zx}
\be\label{Eq:A1-in-e+e}
	A_1 = 1 + \cos(2\phi_1)\;
	\frac{\la\sin^2\theta_2\ra}{\la 1+\cos^2\theta_2\ra}\;
	\frac{
	\sum_ae_a^2C_{\rm Gauss}H_1^{\perp(1/2)a}(z_1)H_1^{\perp(1/2)\bar a}(z_2)}
	{\sum_ae_a^2D_1^a(z_1)D_1^{\bar a}(z_2)}\;.
\ee
Here the average over $\theta_2$ is understood in the range of acceptance 
of the BELLE detector \cite{Abe:2005zx}.

Notice that hard gluon radiation also gives rise to a $\cos(2\phi_1)$-dependence as 
do detector dependent effects. These effects, however, are flavour independent 
and --- as long as the coefficient of the $\cos(2\phi_1)$-modulation is not large 
--- one can get rid of them in the following way \cite{Abe:2005zx}. One considers 
the double ratio of $A_1^U$, where both hadrons $h_1h_2$ are pions of unlike sign, 
to $A_1^L$, where $h_1h_2$ are pions of like sign, i.e.
\be\label{Eq:double-ratio}
	\frac{A_1^U}{A_1^L} = 1 + \cos(2\phi_1) \,P_1(z_1,z_2)\;.
\ee
The observable $P_1(z_1,z_2)$ is given by
\ba\label{Eq:P1}
	P_1(z_1,z_2) \equiv 
	\frac{\la\sin^2\theta_2\ra}{\la 1+\cos^2\theta_2\ra}\;C_{\rm Gauss}\,
	\Biggl[\frac{
	 5H_1^{\perp(1/2)\rm fav}(z_1)H_1^{\perp(1/2)\rm fav}(z_2)
	+7H_1^{\perp(1/2)\rm unf}(z_1)H_1^{\perp(1/2)\rm unf}(z_2)}
	{5D_1^{\rm fav}(z_1)D_1^{\rm fav}(z_2)+
	 7D_1^{\rm unf}(z_1)D_1^{\rm unf}(z_2)}  
	\nonumber\\
	-\frac{
	 5H_1^{\perp(1/2)\rm fav}(z_1)H_1^{\perp(1/2)\rm unf}(z_2)
	+5H_1^{\perp(1/2)\rm unf}(z_1)H_1^{\perp(1/2)\rm fav}(z_2)
	+2H_1^{\perp(1/2)\rm unf}(z_1)H_1^{\perp(1/2)\rm unf}(z_2)}
	{5D_1^{\rm fav}(z_1)D_1^{\rm unf}(z_2)
	+5D_1^{\rm unf}(z_1)D_1^{\rm fav}(z_2)
	+2D_1^{\rm unf}(z_1)D_1^{\rm unf}(z_2)}\Biggr] \nonumber\\
\ea
up to higher order terms in the Collins function. The systematic error of 
the double ratio method was estimated to be small \cite{Abe:2005zx}.

%------ BEGIN FIGURE 5: BEST FIT TO BELLE -----------------------------
%
\begin{figure}[b]
\begin{tabular}{cc}
\includegraphics[width=1.75in]{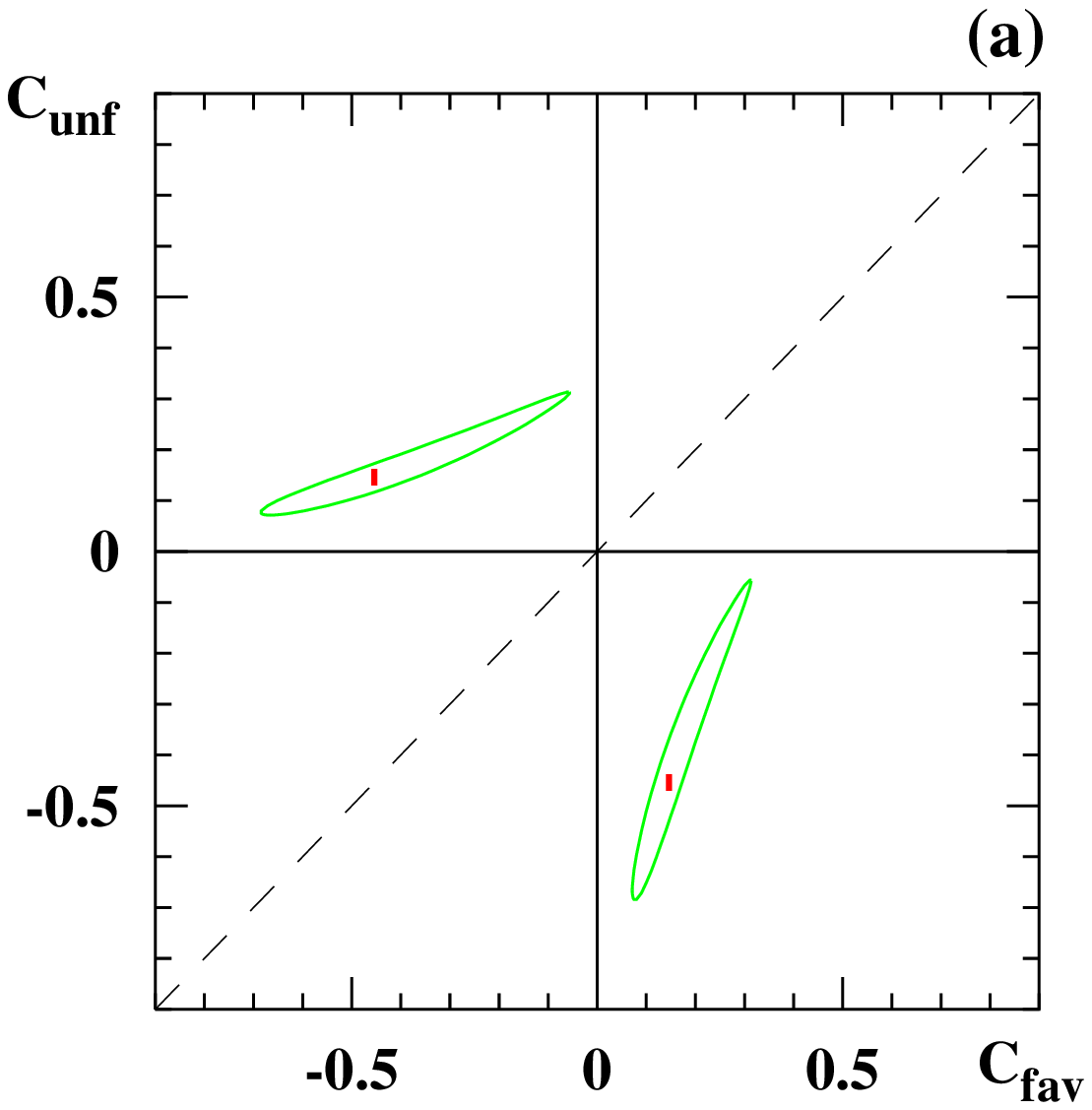}&
\includegraphics[width=1.75in]{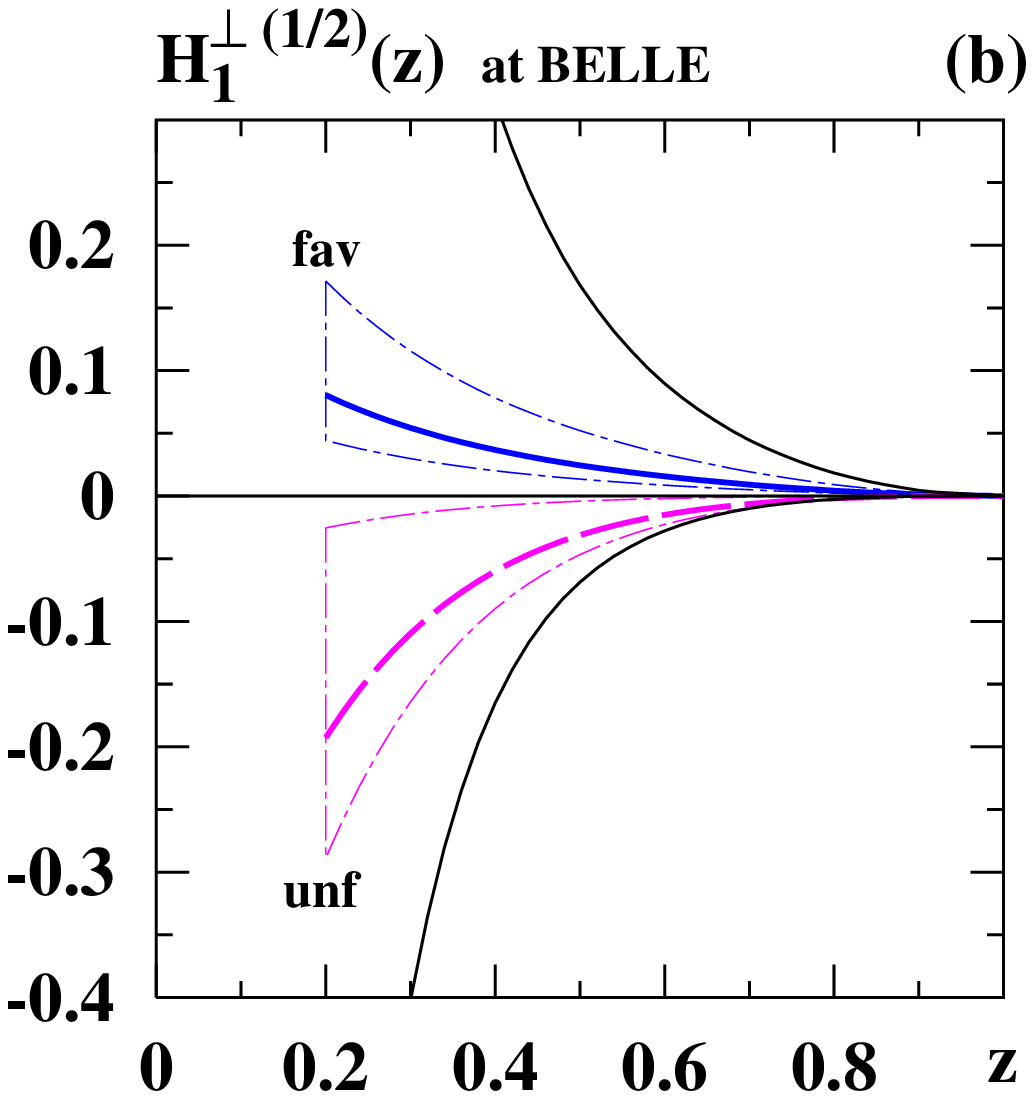}
\end{tabular}
\caption{\label{Fig5:BELLE-best-fit}
	a. The two  best fit solutions for the parameters $C_i$ in 
	the Ansatz (\ref{Eq:ansatz-BELLE}) (indicated as discrete points) 
	and their respective 1-$\sigma$ regions as obtained from a fit
	to the BELLE data \cite{Abe:2005zx}. The solutions
	are symmetric with respect to the line $C_{\rm fav}=C_{\rm unf}$
	indicated by a dashed line.
	b. The best fit for $H_1^{\perp(1/2)a}(z)$ resulting from 
	Fig.~\ref{Fig5:BELLE-best-fit} with the choice $H_1^{\perp\rm fav}>0$
	and $H_1^{\perp\rm unf}<0$ as suggested by the analysis of the
	HERMES experiment, see Sec.~\ref{Sec-2:Collins-effect-in-SIDIS}.}
\end{figure}
%
%------ END FIGURE 5 --------------------------------------------------

In Eq.~(\ref{Eq:P1}) it is assumed that the Collins fragmentation of $s$- and 
$\bar s$-quarks into pions is equal to the unfavoured fragmentation function
defined in Eq.~(\ref{Eq:H-fav-unf}). At the high energies of the BELLE experiment
just below the threshold of $b$-quark production this is a reasonable and commonly 
used assumption for $D_1^a$ \cite{Kretzer:2000yf}.
We assume it here to be valid also for the Collins function. ($D_1^{\rm fav}$ and 
$D_1^{\rm unf}$ in (\ref{Eq:P1}) are defined analogously to Eq.~(\ref{Eq:H-fav-unf}).)
Charm contribution does not need to be considered in Eq.~(\ref{Eq:P1}) since the 
BELLE data are corrected for it \cite{Abe:2005zx}. 

In order to obtain a fit to the BELLE data we adopt the LO-parameterization 
\cite{Kretzer:2000yf} for $D_1^a(z)$ at $Q^2=(10.52\,{\rm GeV})^2$ and choose 
the following simple Ansatz
\be\label{Eq:ansatz-BELLE}
	H_1^{\perp(1/2)a}(z) = C_a \,z\,D_1^a(z)\;.
\ee 
The two free parameters $C_{\rm fav}$ and $C_{\rm unf}$ introduced in
(\ref{Eq:ansatz-BELLE}) can be well fitted to the BELLE data \cite{Abe:2005zx}.
We explored also other Ans\"atze proportional, for example, to $z^2D_1^a(z)$, 
$D_1^a(z)(1-z)$ or $zD_1^a(z)(1-z)$ but none of them gave satisfactory solutions.
The best fit has a $\chi^2_{\rm dof}=0.6$ and is demonstrated in
Fig.~\ref{Fig5:BELLE-best-fit}a in the $C_{\rm fav}$-$C_{\rm unf}$-plane. 
Two different, equivalent, best fit solutions exist. The reason for this 
is that the expression for $P_1(z_1,z_2)$ in Eq.~(\ref{Eq:P1}) is symmetric with 
respect to the exchange $C_{\rm fav} \leftrightarrow C_{\rm unf}$ in our
Ansatz, and manifests itself in Fig.~\ref{Fig5:BELLE-best-fit}a where the two 
solutions are mirror images of each other with respect to the axis defined by 
$C_{\rm fav}=C_{\rm unf}$.

What can unambiguously be concluded from Fig.~\ref{Fig5:BELLE-best-fit}a is that 
the BELLE data require the Collins favoured and unfavoured fragmentation functions
to have opposite sign --- as in the HERMES experiment. On the basis our study of the 
Collins effect in SIDIS we are tempted to select the solution with $C_{\rm fav}>0$ 
and $C_{\rm unf}<0$ in Fig.~\ref{Fig5:BELLE-best-fit}a as the appropriate one.
Our result is thus
\be\label{Eq:best-fit-BELLE}
	C_{\rm fav}=0.15\;,\;\;\;C_{\rm unf}=-0.45\;
\ee
and the resulting best fits and their 1-$\sigma$ regions are shown in 
Fig.~\ref{Fig5:BELLE-best-fit}b for $z>0.2$ which is the low-$z$ cut 
in the BELLE experiment. The results satisfy the positivity condition 
(\ref{Eq:H1perp-1/2mom-and-bound}). Notice that the errors of the
favoured and unfavoured Collins functions are correlated.

%------ BEGIN FIGURE 6: SHOW FIT VS: BELLE DATA -----------------------
%
\begin{figure}
\begin{tabular}{cccc}
\hspace{-0.3cm}
\includegraphics[width=1.75in]{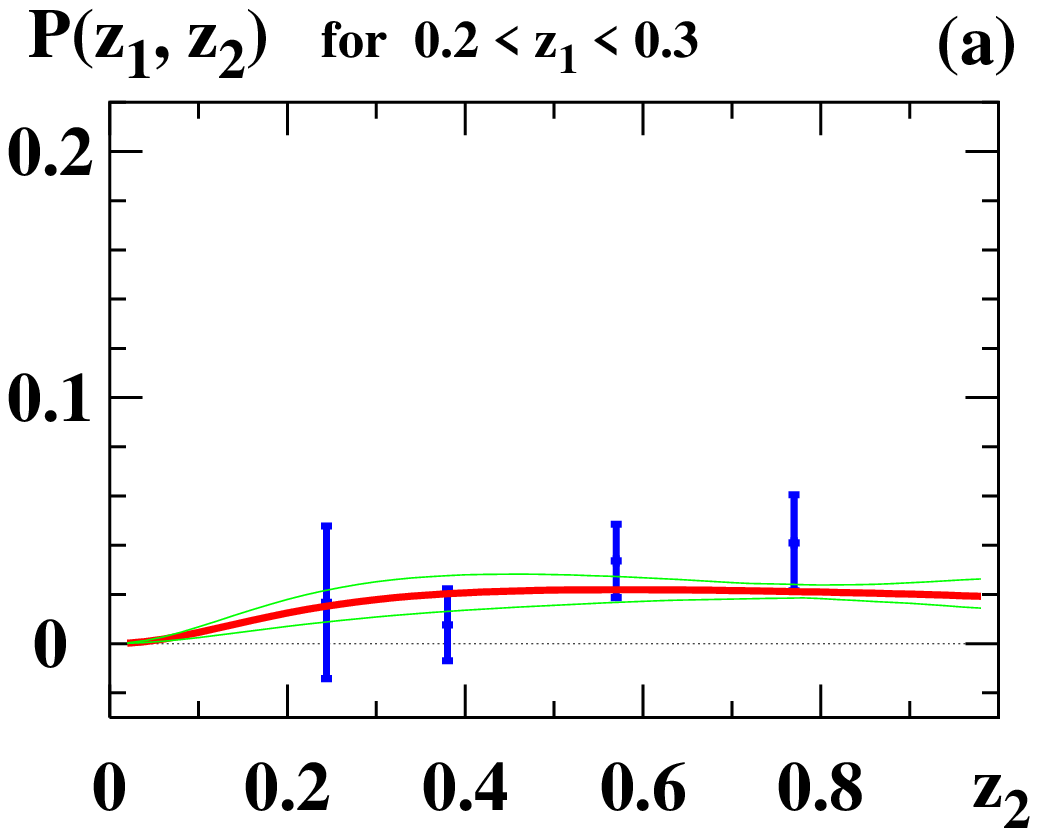}&
\includegraphics[width=1.75in]{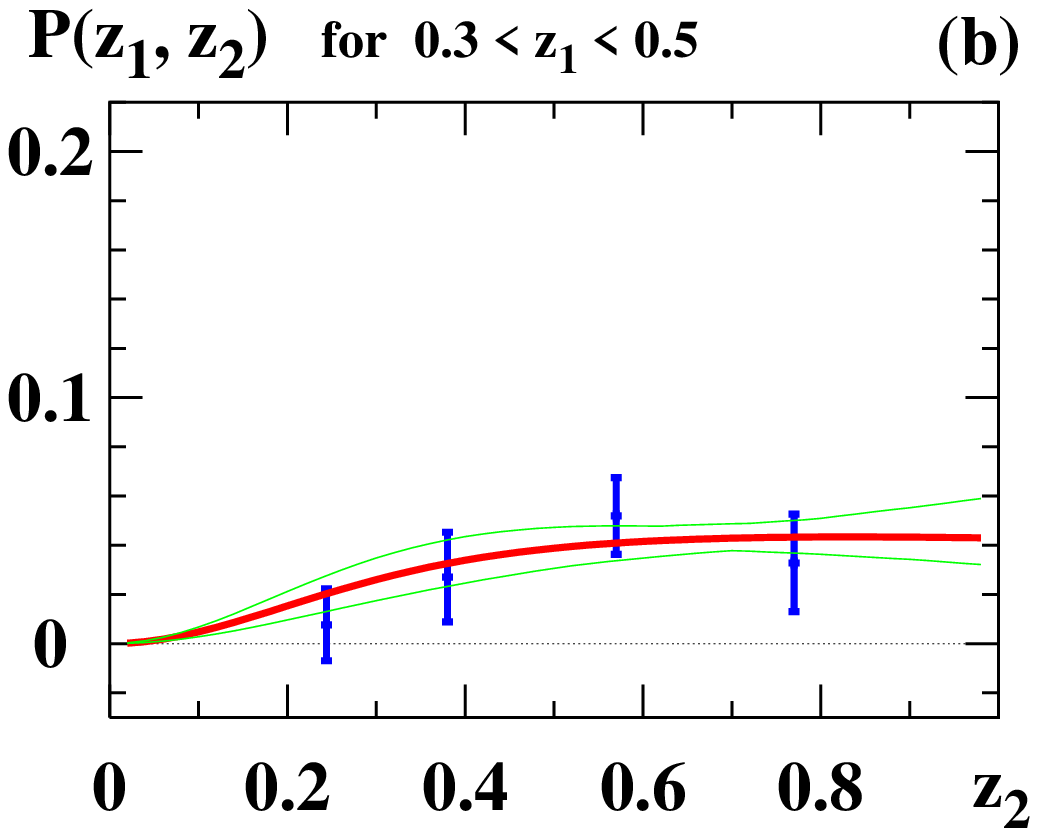}&
\includegraphics[width=1.75in]{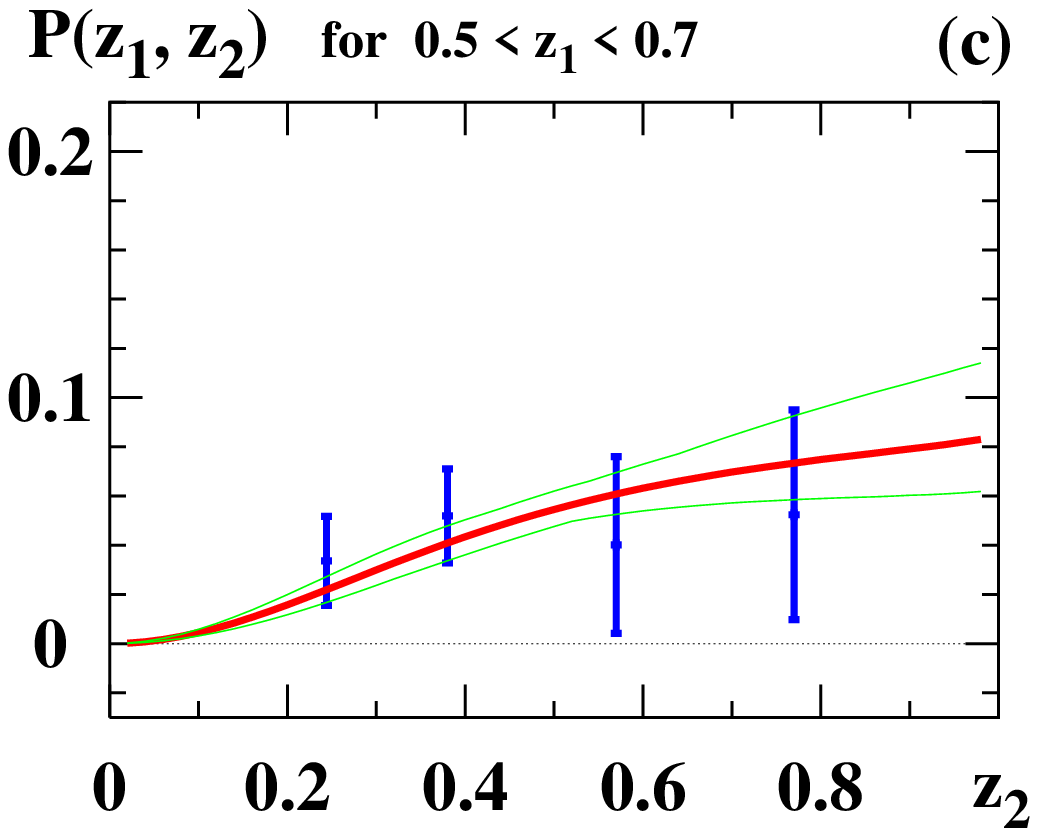}&
\includegraphics[width=1.75in]{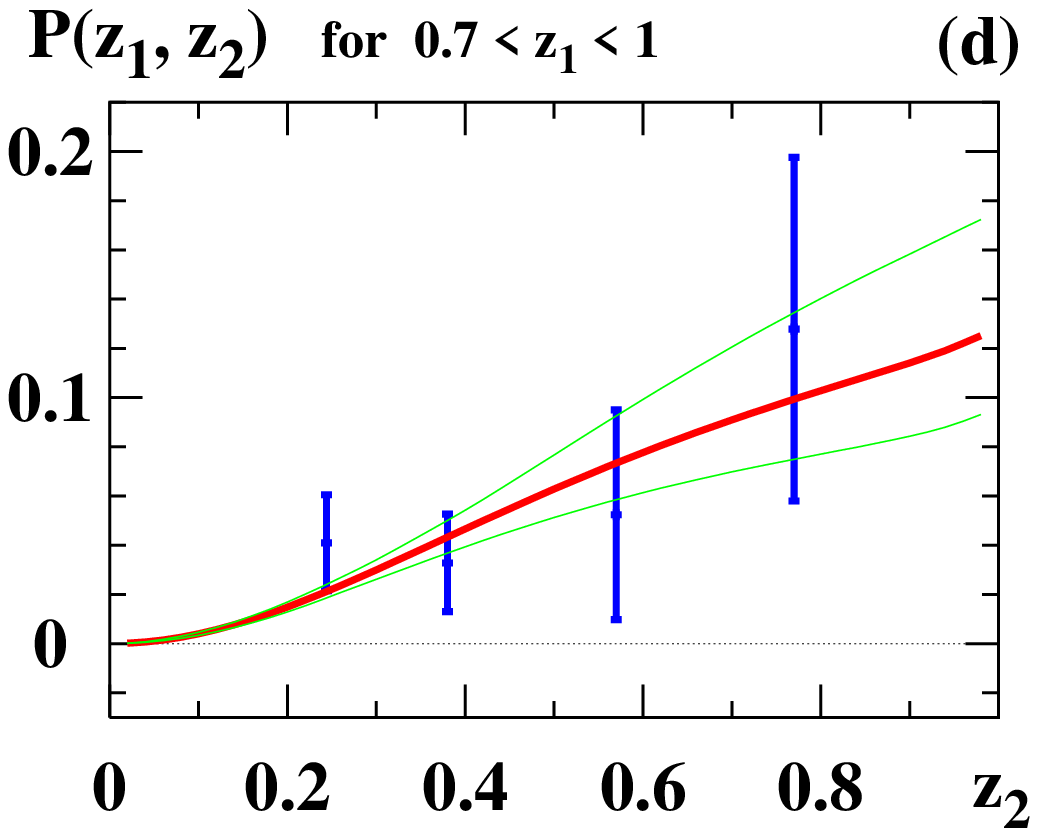}
\end{tabular}
\caption{\label{Fig6:BELLE}
	The observable $P_1(z_1,z_2)$ as defined in 
	Eqs.~(\ref{Eq:double-ratio},~\ref{Eq:P1}) for fixed $z_1$-bins 
	as function of $z_2$. The data are from the BELLE
	experiment \cite{Abe:2005zx}. The theoretical curves are obtained on
	the basis of the fit result shown in Fig.~\ref{Fig5:BELLE-best-fit}.}
\end{figure}
%
%------ END FIGURE 6 --------------------------------------------------

In Fig.~\ref{Fig6:BELLE} the BELLE data \cite{Abe:2005zx}
are compared to the theoretical result for $P_1(z_1,z_2)$ obtained on the 
basis of the best fit shown in Fig.~\ref{Fig5:BELLE-best-fit}b. Here the 
error-correlation of the fit results for the favoured and unfavoured Collins 
function is taken into account and the resulting 1-$\sigma$ error band is 
more narrow than in Fig.~\ref{Fig5:BELLE-best-fit}b.
The description of the BELLE data is satisfactory --- as can be seen from
Fig.~\ref{Fig6:BELLE}.
 (Notice that $P_1(z_1,z_2)$ is symmetric with respect to the exchange
 $z_1\leftrightarrow z_2$ such that the bins with $z_1\neq z_2$ can
 be combined --- as was done in the BELLE analysis \cite{Abe:2005zx}
 and in our fit procedure. Here, for a better overview, 
 these bins are presented separately --- whereby we disregard that
 strictly speaking the statistical error bars for bins with $z_1\neq z_2$
 should then be multiplied by $\sqrt{2}$.)

%===================  SECTION 4: BELLE vs HERMES =====================
\section{Are BELLE and HERMES data compatible?}
\label{SecX:z-dependence-of-SIDIS-data}

In order to compare the Collins effect in SIDIS at HERMES 
\cite{Airapetian:2004tw,Diefenthaler:2005gx} and in 
$e^+e^-$-annihilation at BELLE\cite{Abe:2005zx} it is, strictly speaking, 
necessary to take into account the evolution properties of $H_1^\perp$.
However, in order to get a first rough idea --- which is sufficient at the
present stage --- we instead consider ratios of $H_1^\perp$ to $D_1^a$,
which may be expected to be less scale dependent. For example, integrating 
the BELLE fit result in Fig.~\ref{Fig5:BELLE-best-fit}b over the range of 
HERMES $z$-cuts $0.2<z<0.7$, we obtain the following analyzing powers
\be\label{Eq:Apower-BELLE}
	\frac{\la 2H_1^{\perp(1/2)\rm fav}\ra}{\la D_1^{\rm fav}\ra}
	\biggr|_{\rm BELLE}\;
	= (5.3\;\dots \;20.4)\% \;,\;\;\;
	\frac{\la 2H_1^{\perp(1/2)\rm unf}\ra}{\la D_1^{\rm unf}\ra}
	\biggr|_{\rm BELLE}\;
	= -\,(3.7\;\dots\;41.4)\%  \;.
\ee
Comparing the above numbers (again the errors are correlated) to the result in 
Eq.~(\ref{Eq:Apower-HERMES}) we see that the effects at HERMES and at BELLE 
--- as quantified in Eqs.~(\ref{Eq:Apower-HERMES},~\ref{Eq:Apower-BELLE}) 
--- are of comparable magnitude. 
The central values of the BELLE analyzing powers seem to be systematically 
larger than the HERMES ones. This could partly be attributed to evolution 
effects. However, notice that in the HERMES result (\ref{Eq:Apower-HERMES}) 
in addition the factor $B_{\rm Gauss}<1$ enters, which tends to decrease 
the result.
Thus, the HERMES \cite{Airapetian:2004tw,Diefenthaler:2005gx} and 
BELLE \cite{Abe:2005zx} data seem in good agreement.

\begin{figure}[b]
\begin{tabular}{cc}
\includegraphics[width=1.75in]{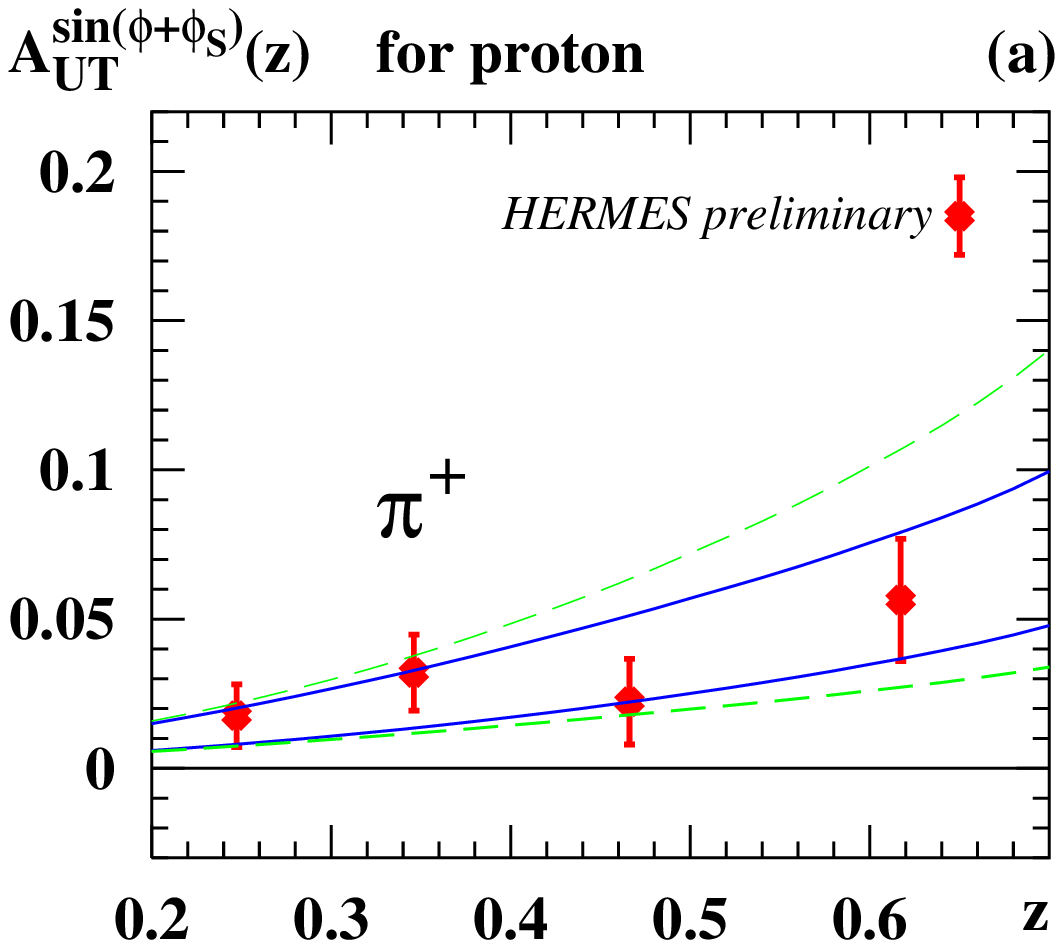}&
\includegraphics[width=1.75in]{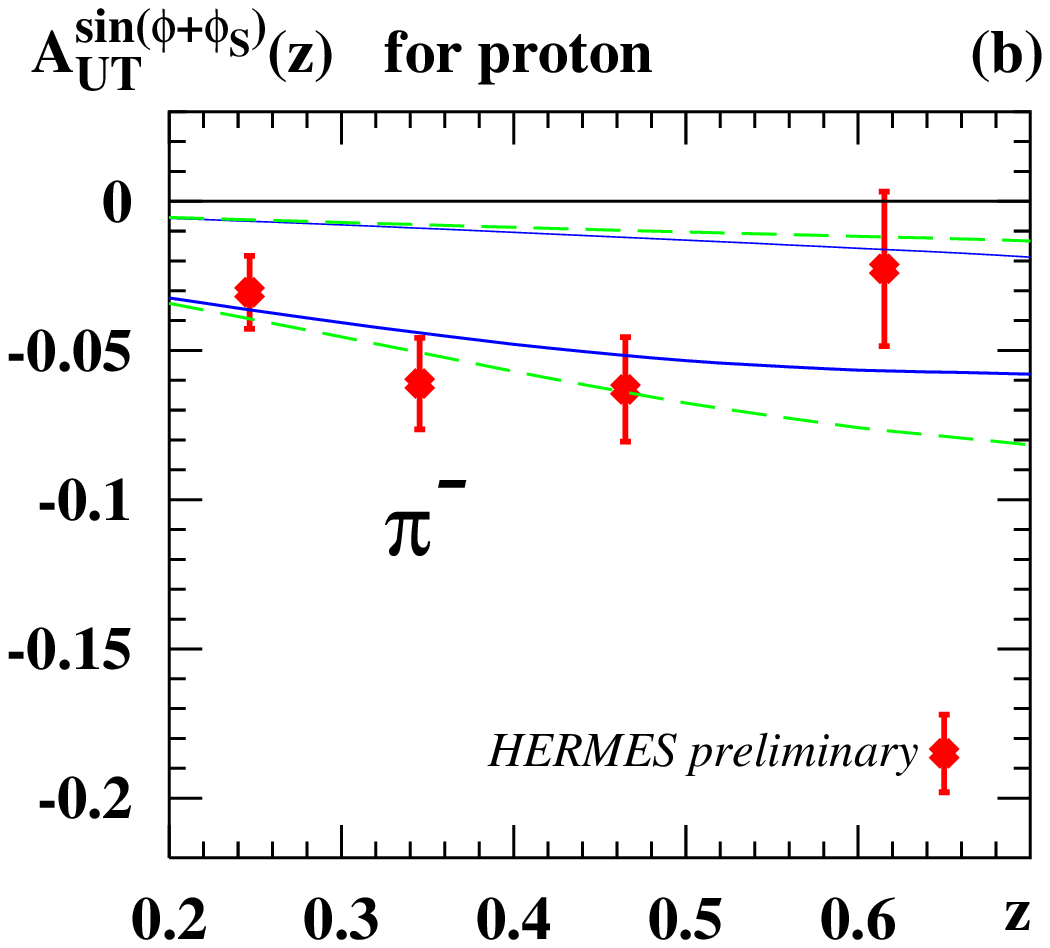}
\end{tabular}
\caption{\label{Fig7:HERMES-AUT-z-from-BELLE}
	The Collins SSA $A_{UT}^{\sin(\phi+\phi_S)}(z)$ as function of $z$.
	The preliminary HERMES data are from \cite{Diefenthaler:2005gx}.
	The theoretical curves are based on the fit of $H_1^\perp$ to 
	the BELLE data shown in Fig.~\ref{Fig5:BELLE-best-fit}b under 
	the assumption (\ref{Eq:assume-weak-scale-dep}). For $h_1^a$ 
	the model prediction \cite{Schweitzer:2001sr} is used. 
	The dashed lines indicate the sensitivity of the 
	SSA to the unknown ratio of the Gaussian widths of $h_1^a$ 
	and $H_1^\perp$ in the range (\ref{Eq:ratio-Gauss-widths}).
	The description of the preliminary HERMES data 
	\cite{Diefenthaler:2005gx} is satisfactory.}
\end{figure}

Encouraged by this observation let us see whether we can describe the HERMES 
data on $A_{UT}^{\sin(\phi+\phi_S)}(z)$ on the basis of the $z$-dependence of 
$H_1^\perp$ concluded from the BELLE data \cite{Abe:2005zx}.
For that let us assume a weak scale-dependence not only for $z$-averaged 
ratios --- as we did above --- but also 
\be\label{Eq:assume-weak-scale-dep}
	\frac{H_1^{\perp(1/2)a}(z)}{D_1^a(z)}\biggr|_{\rm BELLE \;scale}
	\;\;\approx\;
	\frac{H_1^{\perp(1/2)a}(z)}{D_1^a(z)}\biggr|_{\rm HERMES \;scale}
	\;\;.\ee
Nothing is known about the Gaussian widths of the transversity 
distribution and the Collins function which enter the factor 
$B_{\rm Gauss}$ in Eq.~(\ref{Eq:B_Gauss}). Let us therefore assume
their ratio to be similar --- let us say to within a factor of two ---
to the corresponding ratio of the Gaussian widths of the unpolarized 
$f_1^a(x,{\bf p}^2_T)$ and $D_1^a(z,{\bf K}^2_T)$
in Eq.~(\ref{Eq:Gauss-width-unp}), i.e.\ 
\be\label{Eq:ratio-Gauss-widths}
	1 \lesssim \frac{\la{\bf p}^2_{h_1}\ra}{\la{\bf K}^2_{H_1}\ra}
	\lesssim 4 \;.
\ee
It is gratifying to observe that within the range (\ref{Eq:ratio-Gauss-widths}) 
the factor $B_{\rm Gauss}$ varies moderately between $10\,\%$ 
(at small $z\sim 0.3$) and $25\,\%$ (at large $z\sim 0.6$).
Taking into account the 1-$\sigma$ uncertainty of the BELLE fit shown 
in Fig.~\ref{Fig5:BELLE-best-fit} and the variation in 
(\ref{Eq:ratio-Gauss-widths}) we obtain the result in 
Fig.~\ref{Fig7:HERMES-AUT-z-from-BELLE}.
The description of the preliminary HERMES data \cite{Diefenthaler:2005gx} on 
the $z$-dependence of the Collins SSA obtained in this way is satisfactory.

%------ BEGIN FIGURE 7: HERMES AUT(z) FROM BELLE z-DEPENDENCE ---------
%
%
%------ END FIGURE 7 --------------------------------------------------

The good agreement observed in Fig.~\ref{Fig7:HERMES-AUT-z-from-BELLE}
gives further support to our observation that the HERMES 
\cite{Airapetian:2004tw,Diefenthaler:2005gx} and BELLE \cite{Abe:2005zx} 
data are in good agreement. Furthermore, we are lead to the conclusion that
the assumption of weak scale-dependence (\ref{Eq:assume-weak-scale-dep})
is reasonable --- given the accuracy of the data 
\cite{Airapetian:2004tw,Diefenthaler:2005gx,Abe:2005zx}.

Finally, we recall that in the expression for $A_{UT}^{\sin(\phi+\phi_S)}(z)$ 
certain integrals over $x$ enter which extend down to low $x=0.023$ at HERMES
\cite{Airapetian:2004tw,Diefenthaler:2005gx} where the used predictions 
for $h_1^a(x)$ from the model \cite{Schweitzer:2001sr} are at the edge 
of (if not beyond) their range of applicability.
This prevented us from making a direct quantitative extraction of the 
$z$-dependence of the analyzing power from SIDIS data in 
Sect.~\ref{Sec-2:Collins-effect-in-SIDIS}. 
However, here we used the model results \cite{Schweitzer:2001sr} 
merely for sake of a qualitative comparison. The good agreement in 
Fig.~\ref{Fig7:HERMES-AUT-z-from-BELLE} indicates that the uncertainty
due to model-dependence is within the statistical error bars of data and  
uncertainties of the Gaussian Ansatz as estimated in (\ref{Eq:ratio-Gauss-widths}).

% \newpage
%===================  SECTION 5: BELLE vs DELPHI ======================
\section{Are BELLE data and DELPHI result compatible?}
\label{Sec-5:BELLE-vs-DELPHI}

Another interesting question is whether the BELLE and HERMES data are compatible 
with the results obtained from an analysis of DELPHI data \cite{Efremov:1998vd} where 
azimuthal asymmetries in $e^+e^-\to h\,h^\prime X$ at the $Z_0$-peak were studied. 
At DELPHI $h$, $h^\prime$ were leading charged hadrons in opposite jets.
Also in this process the Collins effect gives rise to a $\cos(2\phi_1)$-asymmetry
analog to (\ref{Eq:azim-distr-e+e-}) but with electric charges replaced 
by electro-weak ones and with modifications due to the finite mass and width of 
$Z_0$ \cite{Boer:1997mf,Boer:1997qn}. The cross-section differential in $\phi_1$ 
can be written as \cite{Boer:1997mf,Boer:1997qn}
\be\label{Eq:Apower-DELPHI}
	\frac{\di\sigma(e^+e^-\to h\,h^\prime X)}{\di\phi_1}=
	P_0\,\biggl(1+\cos(2\phi_1)\;P_2\biggr)
\ee
with 
\be\label{Eq:P2-DELPHI-0}
	P_2=
	\frac{\la\sin^2\theta_2\ra}{\la 1+\cos^2\theta_2\ra}\;
	\frac{\int\di z_1\int\di z_2 \,\sum_{h,h^\prime}\sum_{a,\bar a}C_{\rm Gauss}\,
	c_2^a\,H_1^{\perp(1/2)a/h}(z_1)H_1^{\perp(1/2)\bar a/h^\prime}(z_2)}
	{\int\di z_1\int\di z_2\,\,\sum_{h,h^\prime}\sum_{a,\bar a}
	c_1^a\,D_1^{a/h}(z_1)D_1^{\bar a/h^\prime}(z_2)}\;.
\ee
The $c_i^a$ are defined as (the ``$+$'' sign refers to $i=1$, 
the ``$-$'' sign to $i=2$) 
\be
	c_{1,2}^q=(T_3^q-2e^q\sin^2\Theta_{\rm W})^2\pm(T_3^q)^2
\ee
with the weak-isospin $T_3^u=T_3^c=\frac12$, $T_3^d=T_3^s=T_3^b=-\frac12$.

The result obtained from the analysis of the DELPHI data reads \cite{Efremov:1998vd} 
\be\label{Eq:P2-DELPHI-1}
	P_{2,\rm DELPHI} = -(0.26\pm0.18)\% \, .
\ee
In order to have an idea whether the result (\ref{Eq:P2-DELPHI-1}) 
is compatible with BELLE let us make the following rough estimate.
We assume a weak-scale dependence similarly to (\ref{Eq:assume-weak-scale-dep})
and take the charged hadrons observed at DELPHI to be pions and kaons 
(which were, in fact, the most prolific particles).
Then we obtain from the best fit to BELLE data with the unpolarized fragmentation 
functions at $Q^2=M_Z^2$ from \cite{Kretzer:2000yf}
\be\label{Eq:P2-from-BELLE}
	P_{2,\;\rm from\;BELLE} 
	\approx  -(
	 \underbrace{0.052}_{u,d,s\to \pi,\pi}
	+\underbrace{0.007}_{u,d,s\to \pi,K}
	+\underbrace{0.004}_{u,d,s\to K,K}
	+\underbrace{0.103}_{c,b  \to \pi,\pi}
	+\underbrace{0.099}_{c,b  \to \pi,K}
	+\underbrace{0.024}_{c,b  \to K,K})\%
	=	    -(0.06\;\dots\;0.29)\%\;.
\ee
% uds -> pi/pi	-0.00051844
% uds -> pi/ka  -0.00007185
% uds -> ka/ka  -0.00003617
%
% cb -> pi/pi	-0.00103412
% cb -> pi/ka	-0.00099041
% cb -> ka/ka	-0.00024135
The different contributions are explained as follows.
The contribution ``$u,d,s\to \pi,\pi$'' is due to Collins fragmentation 
of light quarks into pions. This is the only contribution which is determined 
by the best fit to the BELLE data without additional assumptions. Since pions and 
kaons are considered as Goldstone bosons of spontaneous chiral symmetry breaking 
one is lead to the conclusion that in the chiral limit \cite{Efremov:2001ia}
\be\label{Eq:chiral-limit}
	\lim\limits_{m_\pi\to 0}\,H_1^{\perp(1/2)a/\pi}(z) =
	\lim\limits_{m_K  \to 0}\,H_1^{\perp(1/2)a/K  }(z) \;.
\ee
Notice that the analog relations with $H_1^\perp$ or $H_1^{\perp(1)}$ would contain 
explicit factors of hadron masses. In the real world the unpolarized fragmentation 
functions for pions and kaons are found to be far off the analog chiral limit
relation. Thus, one may expect the real-world Collins functions to be similarly 
far off the chiral limit relation (\ref{Eq:chiral-limit}). 
However, if the way off the chiral limit to the real world proceeded in a 
spin-independent way, one still could expect the relations 
(\ref{Eq:ansatz-BELLE},~\ref{Eq:best-fit-BELLE}) between the Collins and unpolarized 
fragmentation functions for pions to be valid approximately also for kaons. This is
what we did for the sake of our rough estimate. Future SIDIS and $e^+e^-$ data on kaon
asymmetries will reveal to which extent these assumptions are justified.\footnote{
	The relations (\ref{Eq:ansatz-BELLE},~\ref{Eq:best-fit-BELLE}) cannot 
	hold for all hadrons. Otherwise it would be impossible to satisfy the
	Sch\"afer-Teryaev sum rule \cite{Schafer:1999kn}.}
Finally, to estimate the contribution due to the Collins 
fragmentation of heavier quarks into pions (and kaons) we assume the relations 
(\ref{Eq:ansatz-BELLE}) (and their kaon-analoga) to hold also for charm and bottom 
with $C_c=C_b=C_{\rm unf}$ from (\ref{Eq:best-fit-BELLE}).
These latter assumptions seem natural --- at the $Z^0$-peak, where differences in 
the fragmentation of non-valence quark-flavours $q$ into a specific hadron might 
be expected to be of ${\cal O}(m_q/M_Z)$. In fact, one finds for pions
$D_1^{\rm unf}\equiv D_1^s \approx D_1^{c/\pi}\gtrsim D_1^{b/\pi}$ at $Q^2=M_Z^2$ 
and qualitatively similar for kaons \cite{Kretzer:2000yf}.
But the above assumptions need not to be correct and, for example, the Collins 
fragmentation of heavier flavours could be simply zero. This gives rise to the
uncertainty in the final result in Eq.~(\ref{Eq:P2-from-BELLE}).

The estimate (\ref{Eq:P2-from-BELLE}) --- even if we consider only the Collins
effect of light quarks --- is in reasonable agreement with the result from DELPHI 
in Eq.~(\ref{Eq:P2-DELPHI-1}). 
We refrain from quoting in Eq.~(\ref{Eq:P2-from-BELLE}) the error to 
due to the 1-$\sigma$ uncertainty of the BELLE fit as our result has 
presumably even larger theoretical uncertainties.
Also it is important to keep in mind that the DELPHI result (\ref{Eq:P2-DELPHI-1})
has unestimated (and presumably sizeable) systematic uncertainties.
What is important at this point, however, is that both numbers 
(\ref{Eq:P2-DELPHI-1},~\ref{Eq:P2-from-BELLE}) are of comparable order of
magnitude --- indicating that the DELPHI result \cite{Efremov:1998vd} 
and the BELLE data \cite{Abe:2005zx} could be well due to the same effect.

\section{Remark on the preliminary SMC result}

%------ BEGIN FIGURE 4: Kinematics of e+e- ----------------------------
\begin{wrapfigure}{RD}{6cm}
	\vspace{-0.3cm}
	\centering
        \includegraphics[width=6cm]{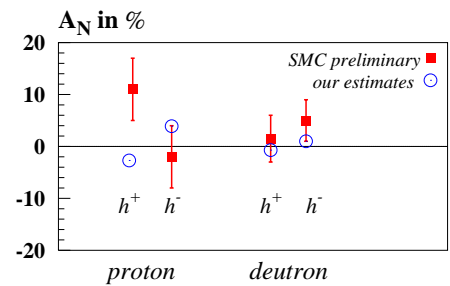}
        \caption{\label{fig8-SMC}
	The Collins SSA $A_N \equiv - A_{UT}^{\sin(\phi+\phi_S)}$ in the
	production of charged hadrons $h^\pm$ in SIDIS from different targets.
	The squares denote the preliminary SMC data \cite{Bravar:1999rq}.
	The open circles show our rough estimates.}
\end{wrapfigure}
%------ END FIGURE 4 -------------------------------------------------
A preliminary result\footnote{
	Notice the different convention used in Ref.~\cite{Bravar:1999rq}
	where the azimuthal distribution of hadrons was analyzed as function
	of the ``Collins angle'' $\phi_C=\phi_h+\phi_S-\pi$. Hence, the opposite
	sign. The same sign convention is used at COMPASS, see 
	caption of Fig.~\ref{Fig3:AUT-x}.}
for the Collins SSA $A_N\equiv - A_{UT}^{\sin(\phi_h+\phi_S)}$ was obtained for 
charged hadrons from an analysis of SMC data on SIDIS from transversely polarized 
proton and deuteron targets at $\la Q^2\ra\sim 5\,{\rm GeV}^2$ and $\la x\ra\sim 0.08$ 
with $\la z\ra\sim 0.45$ and $\la P_{h\perp}\ra \sim (0.5-0.8)\,{\rm GeV}$ 
\cite{Bravar:1999rq}. 
A non-zero Collins SSA for positive hadrons from a proton target was reported, the 
other SSA were found compatible with zero within error bars, see Fig.~\ref{fig8-SMC}.

The charged hadrons at SMC were mainly $\pi^\pm$ \cite{Bravar:1999rq}.
Neglecting the contributions from other hadrons we obtain the results shown
as open circles in Fig.~\ref{fig8-SMC}. Our results are not in conflict with the
preliminary SMC results except for the only case where indications for a non-zero 
SSA were seen, namely for $\pi^+$ from a proton target. Is this a
discrepancy and if so, how could one explain it?

Could, for example, the Collins effect of hadrons other than pions 
be of importance in the SMC kinematics? 
On the basis of Sch\"afer-Teryaev sum rule \cite{Schafer:1999kn} one may, in fact, 
expect cancellations between contributions of different hadrons in SIDIS where the 
Collins effect is linear in $H_1^\perp$.
(In $e^+e^-$-annihilations the effect is proportional to 
$H_1^{\perp q}H_1^{\perp\bar q}$ and different signs of 
the Collins functions for different hadrons cancel.)

It is difficult to address these questions --- in particular because the results 
of Ref.~\cite{Bravar:1999rq} retain a preliminary status and their systematic 
uncertainties are not known. Future data on the Collins effect of other hadrons 
may clarify the situation.

%===================  SECTION 6: VOGELSANG & YUAN =====================
\section{Comparison to the extraction of Ref.~\cite{Vogelsang:2005cs}}
\label{Sec6:compare-to-VY}

In Ref.~\cite{Vogelsang:2005cs} $H_1^{\perp(1/2)}(z)$ was extracted 
from the HERMES data \cite{Diefenthaler:2005gx} in a similar way as
in Sec.~\ref{Sec-2:Collins-effect-in-SIDIS} but with some different 
model assumptions. It is worth to inspect these differences in some 
more detail. This may shed some light on the model dependence of the
two approaches. (Notice the different notation
$\delta\hat{q}^{(1/2)}(z)\equiv - 2H_1^{\perp(1/2)}(z)$ in
\cite{Vogelsang:2005cs}.)

There are two main differences between the approach of 
\cite{Vogelsang:2005cs} and ours. First, in \cite{Vogelsang:2005cs} it was assumed 
that the transverse momentum of the hadrons produced in the SIDIS process is solely
due to the fragmentation process, i.e.\ the possibility was disregarded that partons 
are emitted from the target with  non-zero intrinsic transverse momenta. 
This corresponds to setting $h_1^a(x,{\bf p}_T^2)=h_1^a(x)\,\delta^{(2)}({\bf p}_T)$
and is equivalent to our Gaussian Ansatz (\ref{Eq:Gauss-ansatz}) in the limiting
case ${\bf p}_{h_1}^2\to 0$ in which $B_{\rm Gauss}\to 1$ in 
Eq.~(\ref{Eq:AUT-Collins-1}).
Second, for the transversity distribution the saturation of the Soffer bound 
\cite{Soffer:1994ww} at a low input scale was assumed as discussed in 
\cite{Martin:1997rz}.

The extracted analysing power in SIDIS is valid in any model approach where the 
factorized Ansatz $h_1^a(x,{\bf p}_T^2)=h_1^a(x)\,G({\bf p}_T)$ is assumed (see 
Sec.~\ref{Sec-2:Collins-effect-in-SIDIS}). This was done in \cite{Vogelsang:2005cs} 
and therefore, it is possible to compare our results (\ref{Eq:Apower-HERMES}) 
directly to the fits of Ref.~\cite{Vogelsang:2005cs} which yield
\ba
	    \frac{\la 2H_1^{\perp(1/2)\rm fav}\ra}{\la D_1^{\rm fav}\ra}
	    \biggl|_{\mbox{\footnotesize Ref.~\cite{Vogelsang:2005cs}}}
	&=& \cases{(6.1\pm 0.8)\% & for set I \cr
		   (6.1\pm 0.4)\% & for set II,}\nonumber\\
	    \frac{\la 2H_1^{\perp(1/2)\rm unf}\ra}{\la D_1^{\rm unf}\ra} 
	    \biggl|_{\mbox{\footnotesize Ref.~\cite{Vogelsang:2005cs}}} 
	&=&  \cases{-(11.0\pm 1.3)\% & for set I\cr
	            -(11.5\pm 1.4)\% & for set II.}\label{Eq:Apower-HERMES-VY}
	\ea
We stress that (\ref{Eq:Apower-HERMES}) and (\ref{Eq:Apower-HERMES-VY}) show the 
same quantity but in different models of transverse hadron momenta.
Sets I, II refer to different Ans\"atze in \cite{Vogelsang:2005cs} for the 
Collins fragmentation function. We observe a good agreement with 
(\ref{Eq:Apower-HERMES}). 

We remark that the Ans\"atze of \cite{Vogelsang:2005cs} are of the type 
$H_1^{\perp(1/2)}(z)\propto z(1-z) D_1(z)$. The factor $(1-z)$ in this Ansatz 
is based on general considerations in QCD \cite{Collins:1992kk} and is ``needed'' 
in the approach of \cite{Vogelsang:2005cs} to reproduce the tendency of the 
Collins SSA to decrease (or, at least, not to grow) with increasing $z$ at HERMES 
\cite{Airapetian:2004tw,Diefenthaler:2005gx}. Interestingly we find that such 
an Ansatz is not suited to describe the BELLE \cite{Abe:2005zx} data which extend 
to higher $z$ than the HERMES data \cite{Airapetian:2004tw,Diefenthaler:2005gx}.
In our Gaussian model for transverse parton momenta, where the mean 
transverse momentum of the produced hadrons is taken $z$-independent, 
the ``needed'' decrease of the Collins SSA at larger $z$ is provided by 
the factor $B_{\rm Gauss}$ (\ref{Eq:B_Gauss}).
Thus, although both approaches describe the $z$-dependence of the Collins SSA 
comparably well, the Collins fragmentation functions extracted here and in 
\cite{Vogelsang:2005cs} differ in particular at large-$z$. 
This point nicely illustrates the model dependence both of our results and those 
of Ref.~\cite{Vogelsang:2005cs}. Both approaches are not meant to give a fully 
realistic account of transverse parton momentum dependent process --- but 
are suggested as effective descriptions which are useful at the present stage and
will have to be improved upon the impact of future, more precise data.

Notice that the good agreement of the analyzing powers (\ref{Eq:Apower-HERMES}) 
and (\ref{Eq:Apower-HERMES-VY}) indicates that there is little sensitivity to the 
choice of the model for the transversity distribution function. 
We shall come back to this point below in Sec.~\ref{SecX:what-about-h1}.

%===================  SECTION 7: MODELS ==============================
\section{Comparison to model calculations of the Collins function}

The appearance of any SSA requires the presence of interfering amplitudes 
with different relative phases \cite{Gasiorowich}. The Collins function is 
losely speaking the imaginary part of such interfering amplitudes and this was
explored in model calculations in Ref.~\cite{Collins:1992kk} and further in
\cite{Bacchetta:2001di,Bacchetta:2002tk,Gamberg:2003eg,Bacchetta:2003xn,Amrath:2005gv}.
The common spirit of these calculations is to generate the required imaginary part 
by a one-loop diagram --- using effective quark degrees of freedom, and pions
or (abelian) gluons. Such perturbative calculations do not aim at providing a 
realistic model for the dynamics of the non-perturbative fragmentation process.
Rather their importance is to indicate the existence of the mechanism 
\cite{Metz:2004ya}. It is therefore not surprizing to observe that the models
\cite{Bacchetta:2001di,Bacchetta:2002tk,Gamberg:2003eg,Bacchetta:2003xn,Amrath:2005gv}
have difficulties to reproduce the quantitative properties and in particular the 
flavour dependence of $H_1^\perp$, though this could possibly be achieved by 
exploring the freedom in the choice of model parameters. 
Further studies in this direction are required. 

Presumably better suited to catch features of the non-perturbative fragmentation 
process is the approach of \cite{Artru:1995bh} based on the string fragmentation 
picture. Though quantitatively plagued by uncertainties in the choice and tuning
of parameters this approach {\sl a priori} predicted the favoured and unfavoured 
Collins fragmentation functions to be of similar magnitude and opposite sign.
Explanations of this observation have been given in
\cite{Vogelsang:2005cs} on the basis quark hadron duality considerations.

%\newpage
%===================  SECTION 8: WHAT WE KNOW ABOUT h1 ================
\section{How much do we know about the transversity distribution?}
\label{SecX:what-about-h1}

%------ BEGIN FIGURE 8+9: Apower(r) and picture of h1 -----------------
\begin{wrapfigure}{RD}{6.8cm}
	\centering
	\vspace{-1cm}
        \includegraphics[width=6.8cm]{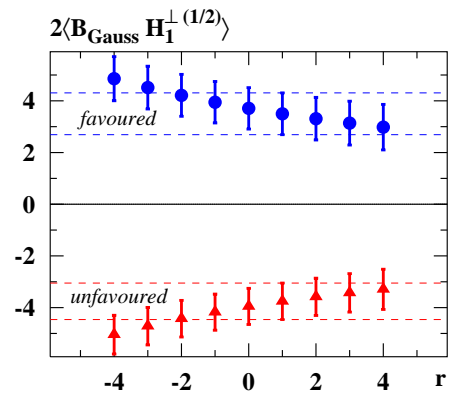}
	\vspace{-0.7cm}
        \caption{\label{figX-Apower-r}
	$\la 2B_{\rm Gauss}H_1^\perp\ra$ as function of the parameter $r$
	defined in Eq.~(\ref{Eq:model-h1d}). 
	The Figure demonstrates that the SIDIS data on the proton Collins SSA 
	\cite{Airapetian:2004tw,Diefenthaler:2005gx} are practically insensitive 
	to $h_1^d(x)$.}
\end{wrapfigure}
After the discussion of the picture of the Collins function emerging 
on the basis of experimental information \cite{Efremov:1998vd,HERMES-new,Airapetian:2004tw,Diefenthaler:2005gx,Alexakhin:2005iw,Abe:2005zx},
it is worthwhile asking how much do we really know about the transversity 
distribution. In other words, to what extent do HERMES and COMPASS data 
\cite{HERMES-new,Airapetian:2004tw,Diefenthaler:2005gx,Alexakhin:2005iw} 
on the Collins SSA in SIDIS constrain $h_1^a(x)$?

In order to gain some insight into this question let us make the following exercise.
We repeat the extraction of the analyzing power from the preliminary HERMES data
\cite{Diefenthaler:2005gx} and use for $h_1^u(x)$ the prediction from the chiral
quark soliton model \cite{Schweitzer:2001sr} but for $h_1^d(x)$ we use instead 
the model
\be\label{Eq:model-h1d}
	h_1^d(x)=r\;h_1^d(x)
	\bigl|_{\mbox{\footnotesize Ref.~\cite{Schweitzer:2001sr}}}\;.\ee
The extracted result for $\la 2B_{\rm Gauss}H_1^\perp\ra$ is then a function of the 
parameter $r$ and its sensitivity to $r$ is shown in Fig.~\ref{figX-Apower-r}. Our 
actual result based on the model \cite{Schweitzer:2001sr} is given for $r=1$ and its
1-$\sigma$ regions are indicated in Fig.~\ref{figX-Apower-r} by dashed lines. 
We make the remarkable observation that the extracted results for the favoured 
and unfavoured $\la 2B_{\rm Gauss}H_1^\perp\ra$ are nearly insensitive to $r$. 
Varying $|h_1^d(x)|$ between zero and four (and more) times the modulus of 
$|h_1^d(x)|$ as predicted by the chiral quark soliton model \cite{Schweitzer:2001sr} 
yields the same results for $\la 2B_{\rm Gauss}H_1^\perp\ra$ within 1-$\sigma$ 
uncertainty. This explains also why our results and those of 
Ref.~\cite{Vogelsang:2005cs} agree, as discussed in Sec.~\ref{Sec6:compare-to-VY}, 
in spite of the different models in particular for the $d$-quark transversity: 
$h_1^d(x)<0$ in our approach \cite{Schweitzer:2001sr} vs.\ $h_1^d(x)>0$ in 
\cite{Vogelsang:2005cs} from saturating the Soffer bound and choosing the positive 
sign. Sensitivity to the smaller (in the model \cite{Schweitzer:2001sr}) 
transversity antiquarks is far weaker.

\begin{wrapfigure}{RD}{8.3cm}
	\centering
        \includegraphics[width=6cm]{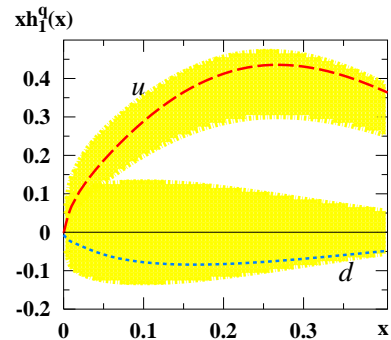}
        \caption{\label{figY-picture-of-h1}
	$xh_1^q(x)$ vs.\  $x$ in the kinematic range of the HERMES experiment.
	The dashed/dotted lines are predictions for $h_1^u/h_1^d$ from
 	the chiral quark soliton model \cite{Schweitzer:2001sr}. 
	The shaded range for the $u$-quark
	is bound from above by the Soffer bound and from below by the estimated 
	theoretical uncertainty of the model result \cite{Schweitzer:2001sr}. 
	The shaded range for the $d$-quark is due to the Soffer bound.
	The Figure demonstrates roughly the emerging picture of the
	transversity distribution.}
\end{wrapfigure}
%
%------ END FIGURE 8+9 -----------------------------------------------
%

Notice that the Soffer bound severely constrains the region of $r$ 
in Fig.~\ref{figX-Apower-r}. This means that the HERMES data 
\cite{Airapetian:2004tw,Diefenthaler:2005gx} leave  $h_1^d(x)$ practically 
unconstrained. The Soffer bound provides basically all we know at present about 
the $d$-quark transversity, see Fig.~\ref{figY-picture-of-h1}.

The situation is different for the $u$-quark transversity. Varying $h_1^u(x)$
within $30\%$ of the chiral quark soliton model prediction (which is within the 
model accuracy) would result in a $30\%$ variation of both the favoured and 
unfavoured $\la 2B_{\rm Gauss}H_1^\perp\ra$. Variations much larger than that
would spoil the good agreement with the BELLE data in 
Sec.~\ref{SecX:z-dependence-of-SIDIS-data}. 
Thus, the model prediction \cite{Schweitzer:2001sr} for $h_1^u(x)$ seems to fit
--- within its theoretical uncertainty and within the uncertainties of our study 
--- precisely in what is needed to explain simultaneously the HERMES 
\cite{Airapetian:2004tw,Diefenthaler:2005gx} and BELLE \cite{Abe:2005zx} data 
with a compatible Collins fragmentation function. This gives rise to the
picture for $h_1^u(x)$ shown in Fig.~\ref{figY-picture-of-h1}.

To conclude the HERMES data \cite{Airapetian:2004tw,Diefenthaler:2005gx} on 
the proton Collins SSA provide constraints for $h_1^u(x)$ but leave $h_1^d(x)$
practically unconstrained. This is not unexpected considering the $u$-quark 
dominance in the proton. In order to learn more about $h_1^d(x)$ and eventually face 
the issue of a flavour separation data from different targets are required. 
The Collins SSA from a deuteron target is smaller than from a proton target and 
more difficult to measure as was shown by COMPASS \cite{Alexakhin:2005iw}. It might 
be promising to explore a $^3$He target, see below Sec.~\ref{Sec-9:predictions}.

%===================  SECTION 9: PREDICTIONS =========================
\section{Predictions for ongoing and future experiments}
\label{Sec-9:predictions}

Further data on transverse target SSA in SIDIS is expected from COMPASS, HERMES 
and the JLab experiments. COMPASS plans to measure with a proton target, and HERMES 
with a deuteron target. Thus, the two experiments will supplement the results 
\cite{HERMES-new,Airapetian:2004tw,Diefenthaler:2005gx,Alexakhin:2005iw} in the
respectively different kinematical regions. The best fit results to the data
\cite{HERMES-new,Airapetian:2004tw,Diefenthaler:2005gx,Alexakhin:2005iw} in 
Figs.~\ref{Fig3:AUT-x} and \ref{Fig7:HERMES-AUT-z-from-BELLE} show basically
the predictions from our study for these experiments for charged pions.

\begin{figure}[t]
\begin{tabular}{ccc}
\includegraphics[width=2.1in]{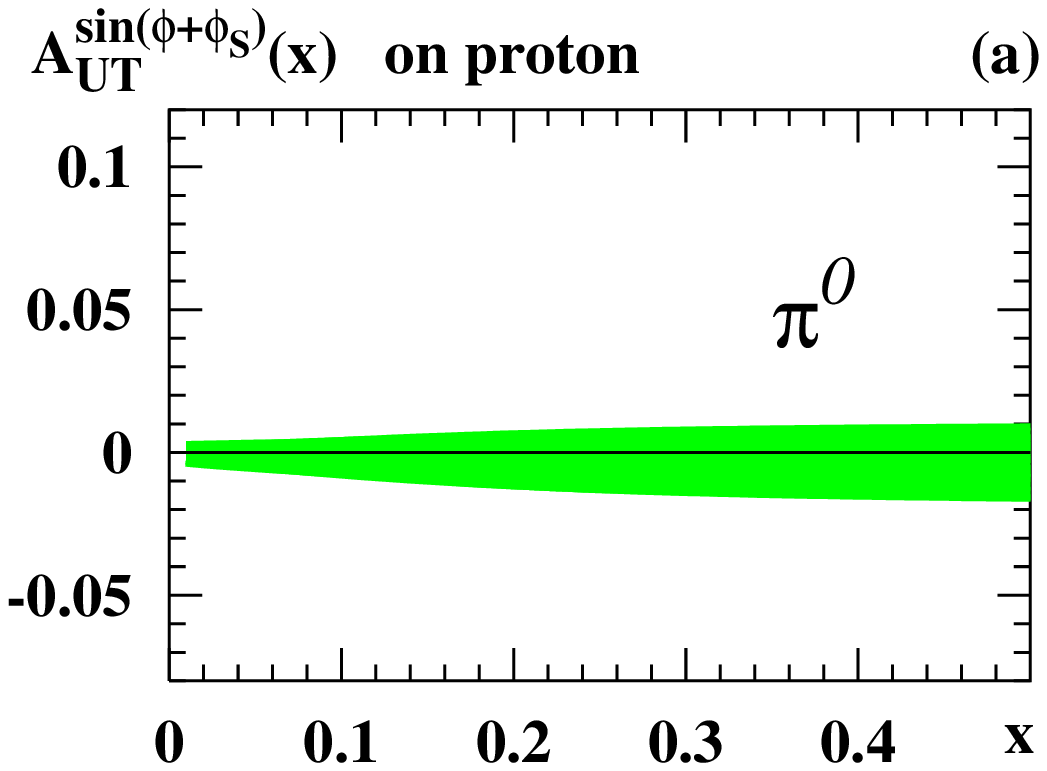}&
\includegraphics[width=2.1in]{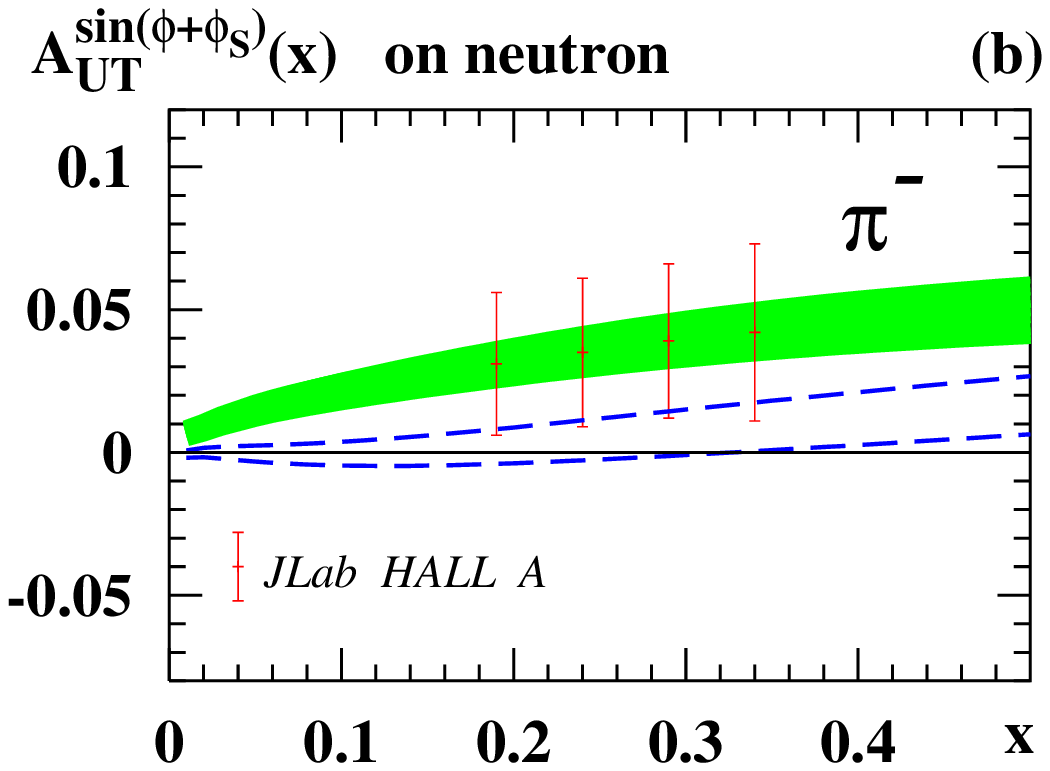}&
\includegraphics[width=2.1in]{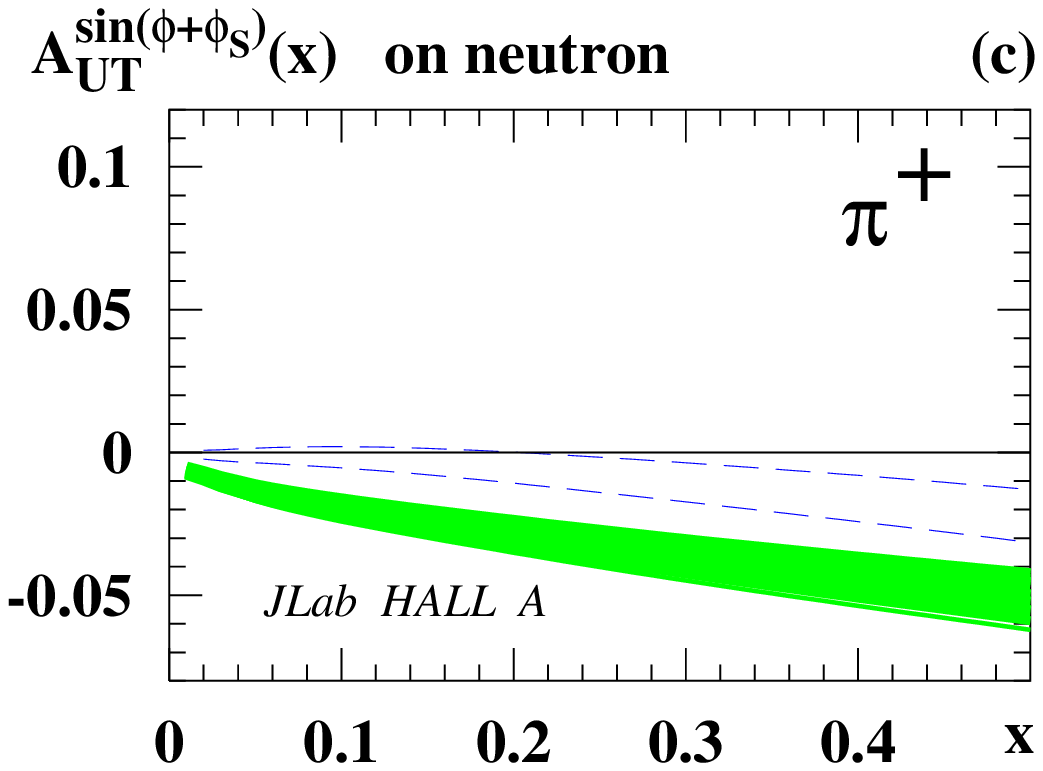}
\end{tabular}
\caption{\label{Fig10:predictions}
	$A_{UT}^{\sin(\phi+\phi_S)}$ vs.\  $x$ as predicted on the basis of our 
	study of the HERMES data in Sect.~\ref{Sec-2:Collins-effect-in-SIDIS} for: 
	(a) $\pi^0$ production from proton, which could 
	be measured at HERMES \cite{HERMES-new} and CLAS \cite{Avakian:2005ps}.
	(b) $\pi^-$ production from a neutron (in practice $^3$He) target, 
	as planned in the HALL-A experiment at JLab. The error bars estimated
	for the HALL-A experiment are taken from Ref.~\cite{Chen:2005dq}.
	The dashed lines indicate tentatively the effects of a different 
	scenario for $h_1^d(x)$, see text.
	(c) The same as (b) but for $\pi^+$.}

\end{figure}

The Collins SSA for neutral pions is an interesting observable. Due to 
isospin symmetry this observable provides information which is already 
contained in the charged pions SSAs. 
However, being an independent experimental information it provides 
a valuable cross check for our understanding of the Collins function.
If our present picture of the  Collins function is correct, one may expect
the $\pi^0$ Collins SSA to be rather small. 
Our prediction for the $\pi^0$ Collins effect on a proton target is shown 
in Fig.~\ref{Fig10:predictions}a. (On a deuteron target the effect is even 
smaller.) Preliminary HERMES data for the $\pi^0$ Collins SSA were shown in 
\cite{HERMES-new} and indicate an effect compatible with zero within large 
uncertainties.

Rather interesting would be data on kaon Collins SSA.
These would not only allow to test whether pions and kaons exhibit 
as Goldstone bosons of chiral symmetry breaking a similar Collins effect, 
see Eq.~(\ref{Eq:chiral-limit}) and the subsequent discussion.
The understanding of kaon Collins fragmentation is required to test
the Sch\"afer-Teryaev sum rule \cite{Schafer:1999kn}. 

The Collins SSA for pions can also be measured in the CLAS and HALL-A 
experiments at JLab on different targets in a somehow different kinematics 
at $Q^2\sim 2\,{\rm GeV}^2$ and for $0.15<x<0.5$ which would provide 
further important information \cite{Avakian:2005ps,Chen:2005dq}. 
For example, the HALL-A collaboration will measure with a transversely 
polarized $^3$He target which --- after nuclear corrections --- will
provide data on the neutron Collins (and Sivers) effect. On the basis of the 
numbers in Eqs.~(\ref{Eq:B-Gauss-H1perp12-fav},~\ref{Eq:B-Gauss-H1perp12-unf})
we estimate for the pion Collins SSA on a ``neutron target'' the results shown 
in Figs.~\ref{Fig10:predictions}b and \ref{Fig10:predictions}c as shaded regions.
The estimates of the SSA could be even more optimistic because in the HALL-A 
experiment larger $z=(0.4-0.6)$ will be probed than at HERMES, and the effect
tends to increase at larger $z$, see Fig.~\ref{Fig7:HERMES-AUT-z-from-BELLE}.
The displayed ``error bars'' in Fig.~\ref{Fig10:predictions}b are projections 
for 24 days of beam time from Ref.~\cite{Chen:2005dq}.

The $\pi^\pm$ Collins SSA on neutron is of similar magnitude as 
the $\pi^\mp$ Collins SSA on proton. However, the pion Collins SSAs on neutron 
are far more sensitive to $h_1^d(x)$.
In order to illustrate this point we present in Figs.~\ref{Fig10:predictions}b
and \ref{Fig10:predictions}c (dotted lines) the result which follows from assuming 
that $h_1^d(x)$ has opposite sign to the chiral quark soliton model prediction 
\cite{Schweitzer:2001sr} (in which $h_1^d(x)$ is negative).
Clearly, on the basis of the HALL-A data it will be possible to discriminate 
the different scenarios, and to constrain $h_1^d(x)$ more strongly.
HALL-A will also be able to measure the Sivers effect for kaons \cite{Chen:2005dq}.

%------ BEGIN FIGURE 10: PREDICTIONS ----------------------------------
%
%
%------ END FIGURE 10 -------------------------------------------------

Before discussing the predictions for azimuthal asymmetries in $e^+e^-$
annihilation (via $\gamma^\ast$) let us first establish how many different 
observables exist in the fragmentation into pions. There are free independent
observables, namely
\ba
	A_1^U(z_1,z_2)
	&=&\;\;	\asym{\pi^+_1}{\pi^-_2}=\asym{\pi^-_1}{\pi^+_2}
	=\left\{\asym{\pi^+_1}{\pi^-_2}+\asym{\pi^-_1}{\pi^+_2}\right\}
	\\
	A_1^L(z_1,z_2)
	&=&\;\;	\asym{\pi^+_1}{\pi^+_2}=\asym{\pi^-_1}{\pi^-_2}
	=\left\{\asym{\pi^+_1}{\pi^+_2}+\asym{\pi^-_1}{\pi^-_2}\right\}
	\\
	A_1^C(z_1,z_2)
	&=&\;\;	\asym{\pi^+_1}{\pi^0_2}=\asym{\pi^-_1}{\pi^0_2}
	 = \;\;	\asym{\pi^0_1}{\pi^+_2}=\asym{\pi^0_1}{\pi^-_2} \nonumber\\
	&=&\!	
	 \left\{\asym{\pi^+_1}{\pi^+_2}+\asym{\pi^-_1}{\pi^-_2}\;+\;
	        \asym{\pi^+_1}{\pi^-_2}+\asym{\pi^-_1}{\pi^+_2}\right\}\;.
	\label{Eq:A_C}\ea
The arrows indicate the opposite jets to which the pions $\pi_i$ belong,
which carry the momentum fractions $z_i$ of the fragmented quarks. (We discard 
the less convenient possibility of combining the events for charged and neutral pions.)
These asymmetries are given by
\ba\label{Eq:observables-at-BELLE}
&&	A_1^U=1+\cos(2\phi_1)\,f_\theta\,\{
	5H_{z_1}^{\rm fav}H_{z_2}^{\rm fav}+5H_{z_1}^{\rm unf}H_{z_2}^{\rm unf}
	+2H_{z_1}^sH_{z_2}^s\}/\{H^a\to D_1^a\}\nonumber\\
&&	A_1^L=1+\cos(2\phi_1)\,f_\theta\,\{
	5H_{z_1}^{\rm fav}H_{z_2}^{\rm unf}+5H_{z_1}^{\rm unf}H_{z_2}^{\rm fav}
	+2H_{z_1}^sH_{z_2}^s\}/\{H^a\to D_1^a\}\nonumber\\
&&	A_1^C=1+\cos(2\phi_1)\,f_\theta\,\{
	5(H_{z_1}^{\rm fav}+H_{z_1}^{\rm unf})(H_{z_2}^{\rm fav}+H_{z_2}^{\rm unf})
	+4H_{z_1}^sH_{z_2}^s\}/\{H^a\to D_1^a\}
\ea
where $f_\theta=\sin^2\theta_2/(1+\cos^2\theta_2)$ and 
$H_z^a\equiv \,C_{\rm Gauss}H_1^{\perp(1/2)a}(z)$ in our approach.

The unlike- and like-sign asymmetries $A_1^U$ and $A_1^L$ were measured at BELLE
(more precisely, the ratio $A_1^U/A_1^L$ was measured) \cite{Abe:2005zx},
see Sect.~\ref{Sec-3:Collins-at-BELLE}.
The third asymmetry $A_1^C$ can be accessed by observing a neutral
and a charged pion as shown in the first line of Eq.~(\ref{Eq:A_C}).
Actually, a linear combination of the observables in the first line of 
Eq.~(\ref{Eq:A_C}) contains the same information. Noteworthy, thanks to isospin 
symmetry the same information is contained in the asymmetry obtained by summing up 
all charged pion events --- as indicated in the second line of Eq.~(\ref{Eq:A_C}). 
Thus, this alternative way allows to access the information content of $A_1^C$ 
possibly more easily --- considering the more difficult $\pi^0$ detection.

To get rid of hard gluon and detector effects, see Sect.~\ref{Sec-3:Collins-at-BELLE},
one may consider the double ratio
\be\label{Eq:BELLE-predict-Pc}
	\frac{A_C}{A_L} = 1 + \cos(2\phi_0)\;P_c(z_1,z_2)\;,
	\ee
where $P_c(z_1,z_2)$ is given to leading order in the analyzing power by
\ba\label{Eq:BELLE-predict-Pc-2}
	P_c(z_1,z_2) = f_\theta\,\biggl[
	\frac{
	5(H_{z_1}^{\rm fav}+H_{z_1}^{\rm unf})(H_{z_2}^{\rm fav}+H_{z_2}^{\rm unf})
	+4H_{z_1}^sH_{z_2}^s}{ \{H^a\to D_1^a\} } \nonumber\\
	- 
	\frac{
	5H_{z_1}^{\rm fav}H_{z_2}^{\rm unf}+5H_{z_1}^{\rm unf}H_{z_2}^{\rm fav}
	+2H_{z_1}^sH_{z_2}^s}{ \{H^a\to D_1^a\} } \biggr]\;
	\ea
in the notation introduced in the context of Eq.~(\ref{Eq:observables-at-BELLE}).

\begin{figure}
\vspace{-0.2cm}
\begin{tabular}{cccc}
\includegraphics[width=1.55in]{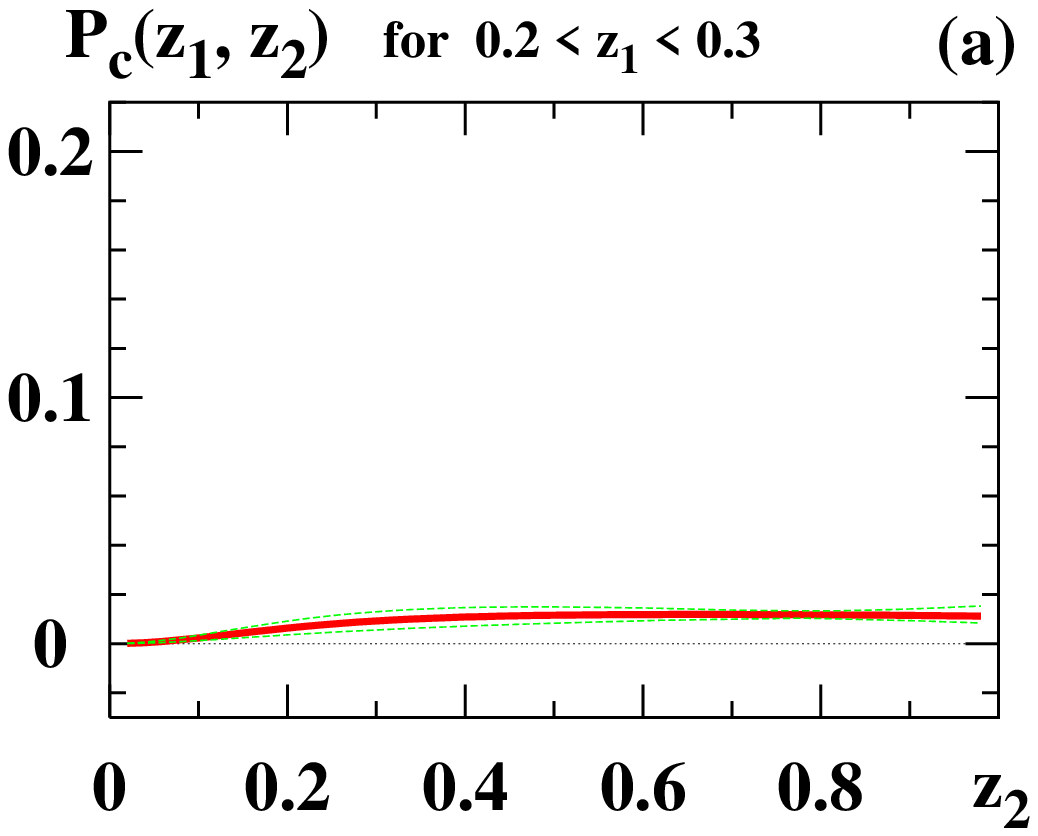}&
\includegraphics[width=1.55in]{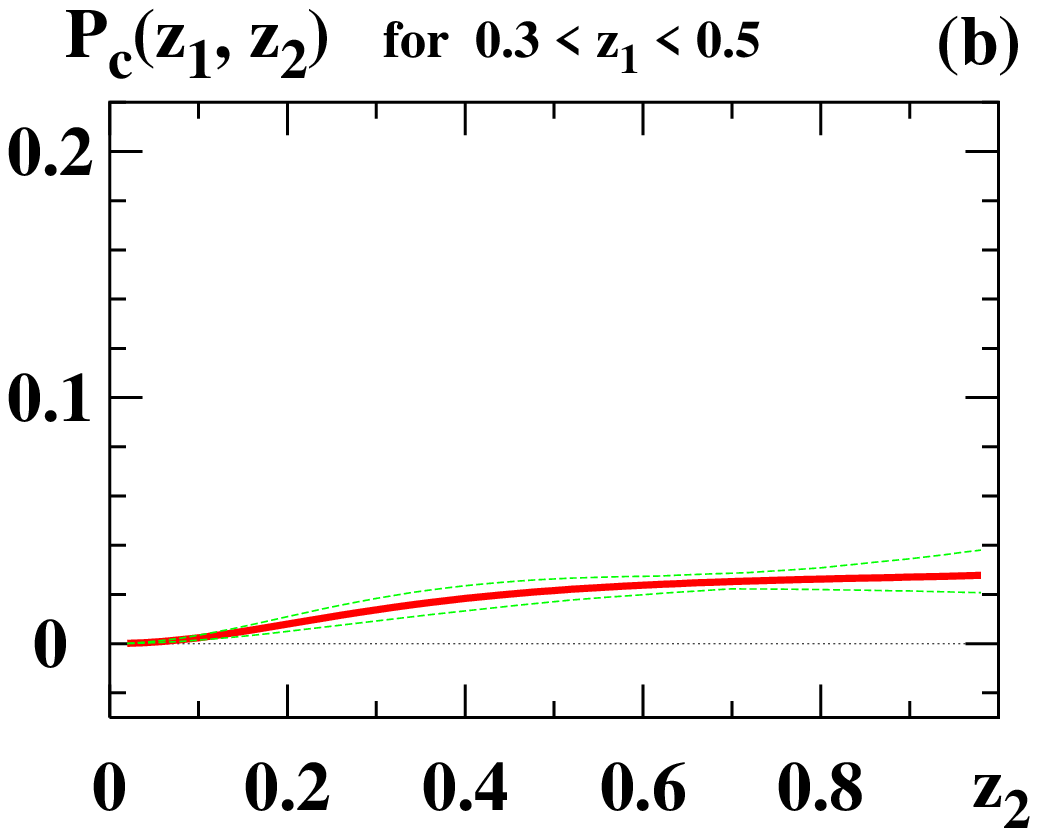}&
\includegraphics[width=1.55in]{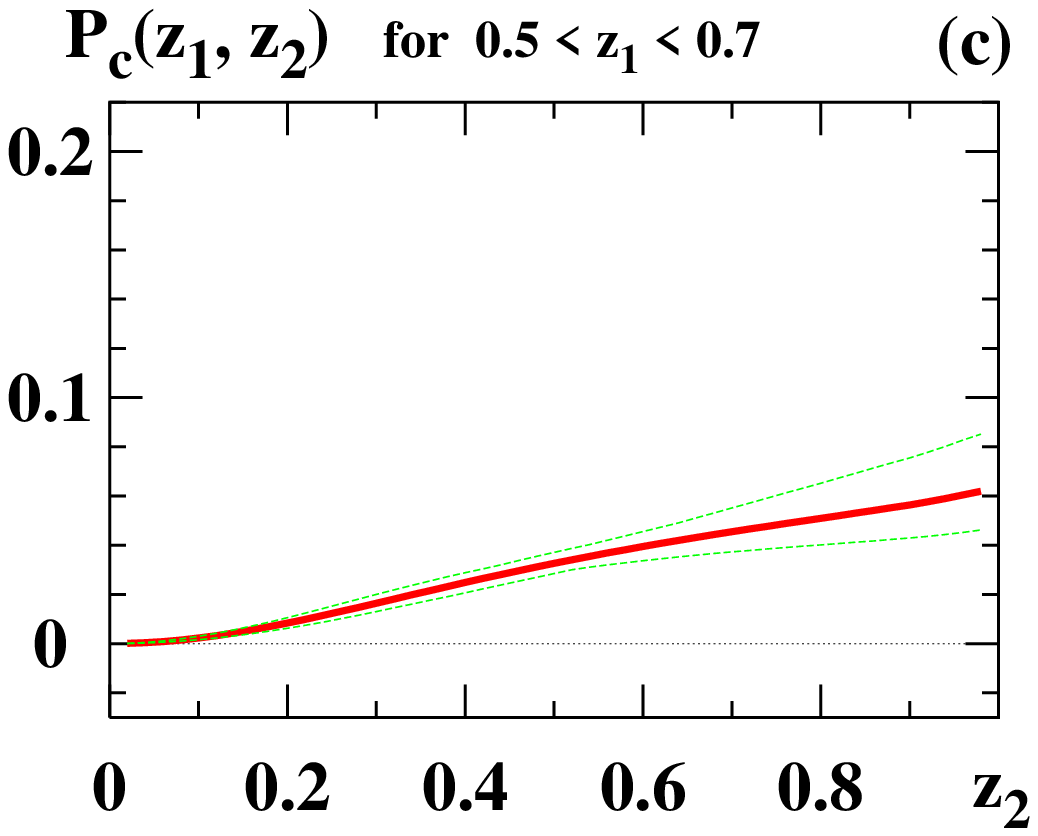}&
\includegraphics[width=1.55in]{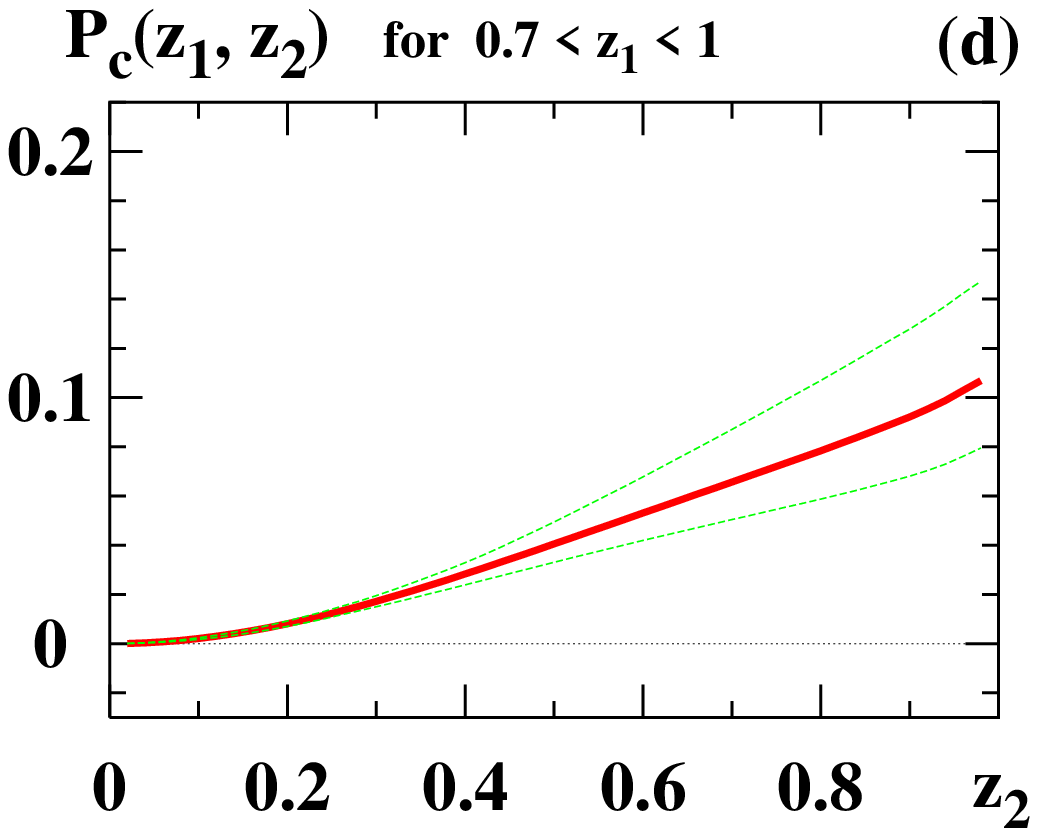}
\end{tabular}
\caption{\label{Fig11:BELLE-predict-Pc}
	The observable $P_c(z_1,z_2)$ as defined in 
	Eqs.~(\ref{Eq:BELLE-predict-Pc}) and (\ref{Eq:BELLE-predict-Pc-2}) 
	for fixed $z_1$-bins as function of $z_2$. The predictions are obtained 
	on the basis of the fit result shown in Fig.~\ref{Fig5:BELLE-best-fit},
	and could be tested in the BELLE experiment.}
\end{figure}

The predictions for $P_c(z_1,z_2)$ made on the basis of our study of the BELLE
data \cite{Abe:2005zx} in Sect.~\ref{Sec-3:Collins-at-BELLE} are shown in
Fig.~\ref{Fig11:BELLE-predict-Pc}. We observe a qualitatively similar picture to
that in Fig.~\ref{Fig6:BELLE} for the observable $P_1(z_1,z_2)$ defined in 
Eqs.~(\ref{Eq:double-ratio},~\ref{Eq:P1}). Thus, $P_c(z_1,z_2)$ could be 
seen in the BELLE experiment in a similarly clear way as $P_1(z_1,z_2)$.
We stress that $P_c(z_1,z_2)$ contains independent experimental information which 
would provide valuable constraints allowing to better pin down the Collins function
and to discriminate between $H_1^{\perp\rm fav}$ and $H_1^{\perp\rm unf}$ using 
less constrained Ans\"atze than e.g.\  (\ref{Eq:ansatz-BELLE}).

%------ BEGIN FIGURE 11: PREDICTIONS FOR BELLE ------------------------
%
%
%------ END FIGURE 11 -------------------------------------------------

%===================  SECTION 10: CONCLUSIONS ========================
\section{Conclusions}

We presented a study of data on azimuthal asymmetries due to the Collins effect 
in SIDIS and in $e^+e^-$ annihilation. Our investigation bears considerable
theoretical uncertainties due to the neglect of soft factors, Sudakov effects,
scale dependence and choice of models for transverse parton momenta and the 
transversity distribution.
We tried to take into account these uncertainties and to minimize them wherever
possible. For example in order to confront results from experiments performed 
at different scales we preferred to compare appropriately defined ratios 
of spin-dependent to spin-independent quantities which might be expected
to be less scale dependent than the respective absolute numbers.

We observed that --- within the uncertainties of our study --- the SIDIS data from
HERMES \cite{HERMES-new,Airapetian:2004tw,Diefenthaler:2005gx} and COMPASS 
\cite{Alexakhin:2005iw} on the Collins SSA from proton and deuteron targets are in 
agreement with each other and with BELLE data \cite{Abe:2005zx} on the azimuthal 
asymmetry in $e^+e^-$ annihilation, 
% NEW:
% and seem compatible with the DELPHI result \cite{Efremov:1998vd} on an 
% azimuthal asymmetry in $e^+e^-$ annihilation at the $Z^0$-peak.
and seem compatible with the indications from DELPHI \cite{Efremov:1998vd} for
an azimuthal asymmetry in $e^+e^-$ annihilation at the $Z^0$-peak.
% END NEW.

The emerging picture for the so far unknown functions is as follows.
\begin{itemize}
\item 	The Collins fragmentation functions show the behaviour 
	$H_1^{\perp(1/2)}\propto zD_1(z)$ at small $z$ as found here and in 
	\cite{Vogelsang:2005cs}.
	Within our model for transverse momenta this relation remains valid also at 
	large $z$, but not in the approach of \cite{Vogelsang:2005cs}.
	The favoured and unfavoured Collins fragmentation functions are of
	comparable magnitude but have opposite sign. SIDIS data require the
	favoured Collins fragmentation to be positive in the convention 
	\cite{Bacchetta:2004jz}.
\item 	The $u$-quark transversity distribution is positive and close 
	(within $30\%$) to saturating the Soffer bound. In contrast, $h_1^d(x)$ 
	and antiquark transversities are entirely unconstrained by present data.
	At the present stage the chiral quark soliton model \cite{Schweitzer:2001sr} 
	provides --- within its theoretical uncertainty --- a useful estimate for 
	$h_1^u(x)$.
\end{itemize}
Future data from SIDIS and $e^+e^-$ colliders will help to refine and improve 
this first picture. Useful for that will be the confrontation of these data with
the estimates for observables in SIDIS and $e^+e^-$ annihilation we presented 
which can be measured in running and planned experiments at HERMES, COMPASS, CLASS,
HALL-A and BELLE.

Particularly helpful would be data on double transverse spin asymmetries in Drell-Yan
lepton pair production from RHIC \cite{Bunce:2000uv} and the planed PAX experiment 
at GSI \cite{PAX} which could provide direct information on $h_1^a(x)$.
In particular, the PAX experiment --- making benefit from a polarized
antiproton beam --- could provide valuable constraints on the $u$-quark transversity 
distribution \cite{PAX-estimates}.
Only the combined analysis of SIDIS, $e^+e^-$ and Drell-Yan data will provide the 
full picture of the Collins fragmentation and transversity distribution function.

\vspace{0.5cm}

\noindent{\bf Acknowledgments.}
We thank Delia Hasch, Harut Avakian, Xiaodong Jiang, Andreas Metz, 
Oleg Teryaev and Feng Yuan for discussions, and the HERMES Collaboration 
for the permission to use the preliminary data \cite{Diefenthaler:2005gx}.
The work is partially supported by BMBF (Verbundforschung), the
COSY-J{\"u}lich project, the Transregio Bonn-Bochum-Giessen, and is
part of the European Integrated Infrastructure Initiative Hadron
Physics project under contract number RII3-CT-2004-506078. 
A.~E.\ is also supported by Grant RFBR 06-02-16215, RF MSE RNP.2.2.2.2.6546
and by the Heisenberg-Landau Program of JINR.

%\newpage
%===================  REFERENCES =====================================

\end{document}